# Data-Driven Uncertainty Quantification and Propagation in Structural Dynamics through a Hierarchical Bayesian Framework

Omid Sedehi[1], Costas Papadimitriou[2], Lambros S. Katafygiotis[3*]

**Abstract**

In the presence of modeling errors, the mainstream Bayesian methods seldom give a realistic account of uncertainties as they commonly underestimate the inherent variability of parameters. This problem is not due to any misconceptions in the Bayesian framework since it is absolutely robust with respect to the modeling assumptions and the observed data. Rather, this issue has deep roots in users' inability to develop an appropriate class of probabilistic models. This paper bridges this significant gap, introducing a novel Bayesian hierarchical setting, which breaks time-history vibrational responses into several segments so as to capture and identify the variability of inferred parameters over multiple segments. Since computation of the posterior distributions in hierarchical models is expensive and cumbersome, novel marginalization strategies, asymptotic approximations, and maximum a posteriori estimations are proposed and outlined under a computational algorithm aiming to handle both uncertainty quantification and propagation tasks. For the first time, the connection between the ensemble covariance matrix and hyper distribution parameters is characterized through approximate estimations. Experimental and numerical examples are employed to illustrate the efficacy and efficiency of the proposed method. It is observed that, when the segments correspond to various system conditions and input characteristics, the proposed method delivers robust parametric uncertainties with respect to unknown phenomena such as ambient conditions, input characteristics, and environmental factors.

[1] Ph.D. Candidate, Department of Civil and Environmental Engineering, The Hong Kong University of Science and Technology, Hong Kong, China; Department of Civil Engineering, Sharif University of Technology, Tehran, Iran, osedehi@connect.ust.hk
[2] Professor, Department of Mechanical Engineering, University of Thessaly, Volos, Greece, costasp@uth.gr
[3*] Professor, Department of Civil and Environmental Engineering, The Hong Kong University of Science and Technology, Hong Kong, China, katafygiotis.lambros@gmail.com (Corresponding Author)

## 1. Introduction

Updating dynamical models based on vibrational data has received growing interest over the last two decades [1–3]. Since such problems are exposed to extensive sources of uncertainty attributed to modeling and measurement errors, using probabilistic methods have become inevitable. Bayesian statistical framework embeds structural models within a class of probability model for describing the model uncertainty in accordance with the misfit between the model and measured outputs [4]. This probability model referred to as likelihood function is combined with a prior probability distribution through the Bayes' rule so as to compute posterior distribution of the model parameters [1,5]. The posterior distribution can then be employed to compute a posterior predictive distribution for response quantities of interest (QoI) [6,7]. When the posterior distribution is concentrated around a sharp peak, the model is called globally identifiable with respect to the observed data [8]. At the same time, the model is regarded as locally identifiable or unidentifiable when multiple or no peak is identified, respectively [9,10]. In practice, it is commonly preferred to construct identifiable models since it allows applying an efficient asymptotic approximation suggested in [1,11]. The uncertainty identified using this framework can be attributed to the lack of knowledge about the actual values of the parameters, which is often reducible as new observations are incorporated. This interpretation of uncertainty remains valid as long as the modeling errors do not induce considerable variabilities in the inferred parameters [12]. Nevertheless, dynamical models are highly misspecified with respect to environmental parameters, ambient conditions, and input characteristics [13]. Consequently, the identified parameters demonstrate predominant variability when they are inferred from multiple data sets. Furthermore, this variability cannot be treated as reducible when new observations are discovered, as opposed to the uncertainty computed using the non-hierarchical Bayesian statistical framework [12,14]. Treatment of these two different sources of uncertainty has been a fundamental research challenge for years and sustained at the forefront of Bayesian criticisms.

The mainstream Bayesian model updating methods [4] are also based on stationary assumptions of the prediction errors statistics. However, prediction errors are often highly non-stationary and can vary depending on the input characteristics, modeling errors, and measurement noise. For slow-varying time-history stochastic processes, a general methodology is to capture the non-stationary effects using windows with limited lengths that moves slowly such that within each window the stationary assumptions are expected to be valid [15–17]. Nevertheless, introducing this concept to Bayesian time-domain model updating methods is often challenging due to the difficulties in fusing the information from different segments.

Multilevel Bayesian methods allow using more flexible assumptions regarding both the model parameters and prediction error probability distributions [18]. Development of hierarchical Bayesian models has brought about many successful applications in different scientific disciplines [18,19]. In molecular dynamics, hierarchical models have recently been developed for calibrating parametric models and fusing heterogeneous experimental data from different system operating conditions [20–22]. In structural dynamics, Behmanesh et al. [14] have developed a hierarchical framework to model and consider the variability of modal parameters over dissimilar experiments. This framework has found extensive applications in uncertainty quantification and propagation of dynamical models based on experimental modal data when they are updated and calibrated under modeling errors [23–25]. Nagel and Sudret [26,27] have proposed a unified multilevel Bayesian framework for calibrating dynamical models for the special case of having noise-free vibration measurements.

Hierarchical models can promote sparsity in Bayesian estimations when implemented through automatic relevance determination method [28,29]. Development of sparse hierarchical models for structural damage identification has led to successful applications as well [30–32]. Sedehi et al. [12] have developed a novel hierarchical Bayesian framework for time-domain model updating and response predictions offering numerous enhancement and improvements over the mainstream Bayesian methods. Having inspired by the recent advances in hierarchical modelling techniques, this paper implements a hierarchical setting to break time-history vibrational data into multiple segments aiming to model and capture the non-stationary effects induced due to modeling errors. As using the hierarchical model gives rise to the number of involved parameters, efficient Laplace asymptotic approximations associated with novel marginalization strategies are proposed for both the model inference and response predictions. Compared to [12] where Markov chain Monte Carlo (MCMC) sampling methods are used to compute the marginal posterior distributions, this paper delivers a new method to compute the maximum a posteriori (MAP) estimations of the hyper-parameters. As searching for the MAP estimations can encounter problems such as trapping into local optimum, the algorithm is accompanied by analytical derivatives and approximate estimations. Although the primary motive of proposing the approximate MAP estimations is to enhance the convergence, simplistic interpretation of the hierarchical modeling are offered for the first time. It is also demonstrated how the MAP estimations of the hyper-parameters can be used to propagate the uncertainty for response predictions. In the end, the proposed method is tested and verified through numerical and experimental examples.

This paper continues with the next section explaining fundamental assumptions used for constructing a hierarchical probabilistic model. It is followed by Sections 3 and 4 which represent the mathematical derivations of the proposed framework and end up with a computational algorithm. Two illustrative examples are included in Section 5 for testing and verifying the proposed method. Summary and conclusions appears in the end.

## 2. Probabilistic Hierarchical Model

Let $\mathbf{D} = \{D_i\}_{i=1}^{N_D}$ denote a family of data sets comprising $N_D$ statistically-independent data sets. Each data set $D_i$ encompasses discrete-time vibrational response of a dynamical system corresponding to $N_0$ degrees-of-freedom (DOF) subjected to the known input $U_i = \{\mathbf{U}_i(k\Delta t_i) \in \mathbb{R}^{N_I}, k = 1,...,n_i\}$. Therefore, one can express $D_i$ as

$$D_i = \{\mathbf{Y}_i(k\Delta t_i) \in \mathbb{R}^{N_0}, k = 1,...,n_i\} \tag{1}$$

where $\Delta t_i$ is the sampling interval corresponding to the $i^{\text{th}}$ data set; $n_i$ is the number of samples within the $i^{\text{th}}$ data set; $\mathbf{U}_i(.)$ and $\mathbf{Y}_i(.)$ respectively denote the discrete-time input and output vectors sampled at $\Delta t_i$ intervals. It should be noticed that the full data set can also be created by splitting up a long time-history data set into a number of segments, where each segment comprises a sufficiently large number of samples. In this case, the segments will represent statistically-independent data sets, and the unknown initial conditions at the beginning of each segment are to be identified.

We aim to update a parametric dynamical model in accordance with multiple data sets/segments. Let $M(\Phi)$ be a structural dynamical model parameterized by a set of parameters denoted by $\Phi$. The set $\Phi$ comprises two kinds of unknown vectors represented by $\boldsymbol{\theta} \in \mathbb{R}^{N_\theta}$ and $\boldsymbol{\psi} \in \mathbb{R}^{N_\psi}$, where $\boldsymbol{\theta}$ corresponds to the parameters characterizing the dynamical behavior (e.g. modal parameters) and $\boldsymbol{\psi}$ corresponds to the unknown initial conditions. Given the model parameters and the input loading $U_i$, the model $M(\Phi)$ produces the following time-history response:

$$X_i = \{\mathbf{X}_i(k\Delta t_i; \boldsymbol{\theta}_i, \boldsymbol{\psi}_i) \in \mathbb{R}^{N_{DOF}}, k = 1, ..., n_i\} \tag{2}$$

where $\mathbf{X}_i(.)$ is a vector of model responses at all observed and unobserved DOF; $\boldsymbol{\theta}_i$ and $\boldsymbol{\psi}_i$ denote the parameters corresponding to the $i^{\text{th}}$ data set; $N_{DOF}$ is the total number of DOF. As the dependence on the model $M(\Phi)$ can be realized from the dependence on the model parameters, it is dropped from the formulations only for the sake of brevity. When this model is used to predict the measured output vector $\mathbf{Y}_i(k\Delta t_i)$, it creates the following prediction errors:

$$\boldsymbol{\varepsilon}_i(k\Delta t_i; \boldsymbol{\theta}_i, \boldsymbol{\psi}_i) = \mathbf{Y}_i(k\Delta t_i) - \mathbf{S}_o \mathbf{X}_i(k\Delta t_i; \boldsymbol{\theta}_i, \boldsymbol{\psi}_i) \tag{3}$$

where $\mathbf{S}_o \in \mathbb{R}^{N_0 \times N_{DOF}}$ is a known matrix selecting the observed output quantities from all output QoI, and $\boldsymbol{\varepsilon}_i(.) \in \mathbb{R}^{N_0}$ denotes the prediction errors. In general, the prediction errors are unknown stochastic processes, which can vary depending on the model parameters, the initial conditions, the input loadings, modeling errors, and measurement noise. The relationship between the prediction errors and these factors is seldom known. Therefore, we model them probabilistically by assuming the data points to be statistically independent and identically distributed (*i.i.d.*), described using Gaussian distributions. Note that using Gaussian distributions is optimum in the maximum entropy sense [33]. Moreover, the statistical characteristics of the prediction errors are essentially specific to each data set such that the model can perform either satisfactorily or poorly to estimate the actual response. Given these assumptions, the prediction errors can be described as

$$p\left(\boldsymbol{\varepsilon}_i(k\Delta t_i;\boldsymbol{\theta}_i,\boldsymbol{\psi}_i)\right)=N\left(\boldsymbol{\varepsilon}_i(k\Delta t_i;\boldsymbol{\theta}_i,\boldsymbol{\psi}_i)\,|\,\mathbf{0},\boldsymbol{\Sigma}_i^{\varepsilon}\right) \tag{4}$$

where $N\left(.|.\,,\,.\right)$ denotes a Gaussian distribution, and $\boldsymbol{\Sigma}_i^{\varepsilon}\in\mathbb{R}^{N_0\times N_0}$ is prediction errors covariance matrix expressed as

$$\boldsymbol{\Sigma}_i^{\varepsilon}=\mathrm{diag}\left\{\left(\sigma_{\varepsilon}^{(i,1)}\right)^2 \quad \cdots \quad \left(\sigma_{\varepsilon}^{(i,j)}\right)^2 \quad \cdots \quad \left(\sigma_{\varepsilon}^{(i,N_0)}\right)^2\right\} \tag{5}$$

Here, $\left(\sigma_{\varepsilon}^{(i,j)}\right)^2$ is the variance corresponding to the $i^{\text{th}}$ data set and the $j^{\text{th}}$ DOF. This diagonal matrix follows from the *i.i.d.* data points corresponding to different DOF. In the remainder, we use $P(\Sigma)$ to represent this probability distribution, where its unknown parameters are collected in the set $\Sigma$. Due to the statistical independence of the data points, the individual likelihood function of each data set is described as

$$p\left(D_i\,|\,\boldsymbol{\theta}_i,\boldsymbol{\psi}_i,\boldsymbol{\Sigma}_i^{\varepsilon}\right)=\prod_{k=1}^{n_i}N\left(\boldsymbol{\varepsilon}_i(k\Delta t_i;\boldsymbol{\theta}_i,\boldsymbol{\psi}_i)\,|\,\mathbf{0},\boldsymbol{\Sigma}_i^{\varepsilon}\right)=\prod_{j=1}^{N_0}\prod_{k=1}^{n_i}N\left(\varepsilon_{i,j}(k\Delta t_i;\boldsymbol{\theta}_i,\boldsymbol{\psi}_i)\,|\,0,\left(\sigma_{\varepsilon}^{(i,j)}\right)^2\right) \tag{6}$$

where $\varepsilon_{i,j}(k\Delta t_i;\boldsymbol{\theta}_i,\boldsymbol{\psi}_i)$ is the prediction error corresponding to the $i^{\text{th}}$ data set, the $j^{\text{th}}$ DOF, and the $k^{\text{th}}$ time sample. Likewise, the likelihood function of the full data set is described in accordance with the statistical independence of the data sets (segments) giving:

$$p\left(\mathbf{D}\,|\,\{\boldsymbol{\theta}_i,\boldsymbol{\psi}_i,\boldsymbol{\Sigma}_i^{\varepsilon}\}_{i=1}^{N_D}\right)=\prod_{i=1}^{N_D}p\left(D_i\,|\,\boldsymbol{\theta}_i,\boldsymbol{\psi}_i,\boldsymbol{\Sigma}_i^{\varepsilon}\right)=\prod_{i=1}^{N_D}\prod_{j=1}^{N_0}\prod_{k=1}^{n_i}N\left(\varepsilon_{i,j}(k\Delta t_i;\boldsymbol{\theta}_i,\boldsymbol{\psi}_i)\,|\,0,\left(\sigma_{\varepsilon}^{(i,j)}\right)^2\right) \tag{7}$$

After these preliminaries, the correlation between the data-set-specific parameters should be modeled. In practice, the input characteristics and the prediction errors parameters can be different over dissimilar data sets. However, the dynamical characteristics are expected to share similarities over data sets, provided that the system does not undergo damaging loading scenarios. Therefore, we regard both $\boldsymbol{\psi}_i$'s and $\boldsymbol{\Sigma}_i^{\varepsilon}$'s as statistically independent parameters while $\boldsymbol{\theta}_i$'s are correlated under a hierarchical setting involving a two-level conditional probability model. The upper level is a hyper probability model representing the second-moment statistics of $\boldsymbol{\theta}_i$'s, whereas in the lower level all $\boldsymbol{\theta}_i$'s are positioned as the drawn samples of the hyper distribution. This hierarchy presumes that $\boldsymbol{\theta}_i$'s can vary over different data sets while maintaining the same statistics over dissimilar data sets. The hyper distribution is selected to be a Gaussian distribution with the mean vector $\boldsymbol{\mu}_{\boldsymbol{\theta}} \in \mathbb{R}^{N_\theta}$ and covariance matrix $\boldsymbol{\Sigma}_{\boldsymbol{\theta\theta}} \in \mathbb{R}^{N_\theta \times N_\theta}$. Again, using Gaussian distribution is an optimal choice when the problem is viewed from a maximum entropy perspective [33]. The parameters of the hyper distribution that often referred to as hyper-parameters are subsumed into the set $\phi = \{\boldsymbol{\mu}_{\boldsymbol{\theta}}, \boldsymbol{\Sigma}_{\boldsymbol{\theta\theta}}\}$, and the hyper probability model is denoted by $P(\phi)$.

The combination of the deterministic model $M(\Phi)$, the prediction errors probability model $P(\Sigma)$, and the hierarchical hyper distribution $P(\phi)$ constitutes a class of multilevel probabilistic model denoted by $\mathbf{M}_p(\Theta)$, where $\Theta = \left\{ \{\boldsymbol{\theta}_i, \boldsymbol{\psi}_i, \boldsymbol{\Sigma}_i^{\varepsilon}\}_{i=1}^{N_D}, \boldsymbol{\mu}_{\boldsymbol{\theta}}, \boldsymbol{\Sigma}_{\boldsymbol{\theta\theta}} \right\}$ is the set comprising all parameters. This probabilistic model is shown in Fig. 1, indicating the conditional dependence between the parameters through the arrows. This graphical demonstration allows constructing the joint prior distribution as follows:

$$p\left(\left\{\boldsymbol{\theta}_i, \boldsymbol{\psi}_i, \boldsymbol{\Sigma}_i^{\varepsilon}\right\}_{i=1}^{N_D}, \boldsymbol{\mu}_{\boldsymbol{\theta}}, \boldsymbol{\Sigma}_{\boldsymbol{\theta\theta}}\right) \propto p\left(\boldsymbol{\mu}_{\boldsymbol{\theta}}, \boldsymbol{\Sigma}_{\boldsymbol{\theta\theta}}\right) \prod_{i=1}^{N_D} \left[ p\left(\boldsymbol{\theta}_i \mid \boldsymbol{\mu}_{\boldsymbol{\theta}}, \boldsymbol{\Sigma}_{\boldsymbol{\theta\theta}}\right) p\left(\boldsymbol{\psi}_i\right) p\left(\boldsymbol{\Sigma}_i^{\varepsilon}\right) \right] \tag{8}$$

where $p\left(\boldsymbol{\mu}_{\boldsymbol{\theta}},\boldsymbol{\Sigma}_{\boldsymbol{\theta\theta}}\right)$ is the hyper-parameters' prior distribution, $p\left(\boldsymbol{\theta}_i \mid \boldsymbol{\mu}_{\boldsymbol{\theta}},\boldsymbol{\Sigma}_{\boldsymbol{\theta\theta}}\right)$ is the probability distribution of $\boldsymbol{\theta}_i$'s conditional on the hyper-parameters described using the Gaussian distribution $N\left(\boldsymbol{\theta}_i \mid \boldsymbol{\mu}_{\boldsymbol{\theta}},\boldsymbol{\Sigma}_{\boldsymbol{\theta\theta}}\right)$, $p\left(\boldsymbol{\psi}_i\right)$ is the prior distribution of the initial conditions assumed to be non-informatively uniform, and $p\left(\boldsymbol{\Sigma}_i^{\varepsilon}\right)$ is the prior distribution of the prediction error parameters expressed by the Jeffreys non-informative prior distribution giving [19]:

$$p\left(\boldsymbol{\Sigma}_i^{\varepsilon}\right) \propto \prod_{j=1}^{N_0}\left(\sigma_{\varepsilon}^{(i,j)}\right)^{-2} \tag{9}$$

Thus, the Bayes' rule expresses the joint posterior distribution as

$$p\left(\left\{\boldsymbol{\theta}_i,\boldsymbol{\psi}_i,\boldsymbol{\Sigma}_i^{\varepsilon}\right\}_{i=1}^{N_D},\boldsymbol{\mu}_{\boldsymbol{\theta}},\boldsymbol{\Sigma}_{\boldsymbol{\theta\theta}} \mid \mathbf{D}\right) \propto p\left(\mathbf{D} \mid \left\{\boldsymbol{\theta}_i,\boldsymbol{\psi}_i,\boldsymbol{\Sigma}_i^{\varepsilon}\right\}_{i=1}^{N_D}\right) p\left(\left\{\boldsymbol{\theta}_i,\boldsymbol{\psi}_i,\boldsymbol{\Sigma}_i^{\varepsilon}\right\}_{i=1}^{N_D},\boldsymbol{\mu}_{\boldsymbol{\theta}},\boldsymbol{\Sigma}_{\boldsymbol{\theta\theta}}\right) \tag{10}$$

where $p\left(\left\{\boldsymbol{\theta}_i,\boldsymbol{\psi}_i,\boldsymbol{\Sigma}_i^{\varepsilon}\right\}_{i=1}^{N_D},\boldsymbol{\mu}_{\boldsymbol{\theta}},\boldsymbol{\Sigma}_{\boldsymbol{\theta\theta}}\right)$ is the prior distribution given by Eq. (8), and $p\left(\mathbf{D} \mid \left\{\boldsymbol{\theta}_i,\boldsymbol{\psi}_i,\boldsymbol{\Sigma}_i^{\varepsilon}\right\}_{i=1}^{N_D}\right)$ is the likelihood function of the full data set expressed earlier in Eq. (7). Combining Eqs. (7-10) eventually leads to the following joint posterior distribution:

$$p\left(\left\{\boldsymbol{\theta}_i,\boldsymbol{\psi}_i,\boldsymbol{\Sigma}_i^{\varepsilon}\right\}_{i=1}^{N_D},\boldsymbol{\mu}_{\boldsymbol{\theta}},\boldsymbol{\Sigma}_{\boldsymbol{\theta\theta}} \mid \mathbf{D}\right) \propto p\left(\boldsymbol{\mu}_{\boldsymbol{\theta}},\boldsymbol{\Sigma}_{\boldsymbol{\theta\theta}}\right)\prod_{i=1}^{N_D} p\left(D_i \mid \boldsymbol{\theta}_i,\boldsymbol{\psi}_i,\boldsymbol{\Sigma}_i^{\varepsilon}\right) p\left(\boldsymbol{\Sigma}_i^{\varepsilon}\right) N\left(\boldsymbol{\theta}_i \mid \boldsymbol{\mu}_{\boldsymbol{\theta}},\boldsymbol{\Sigma}_{\boldsymbol{\theta\theta}}\right) \tag{11}$$

Due to the large number of involved parameters and the complex structure of this distribution, it is inefficient and cumbersome to calculate the MAP estimations directly from this formulation. In the next section, a novel marginalization scheme is proposed to simplify the uncertainty quantification.

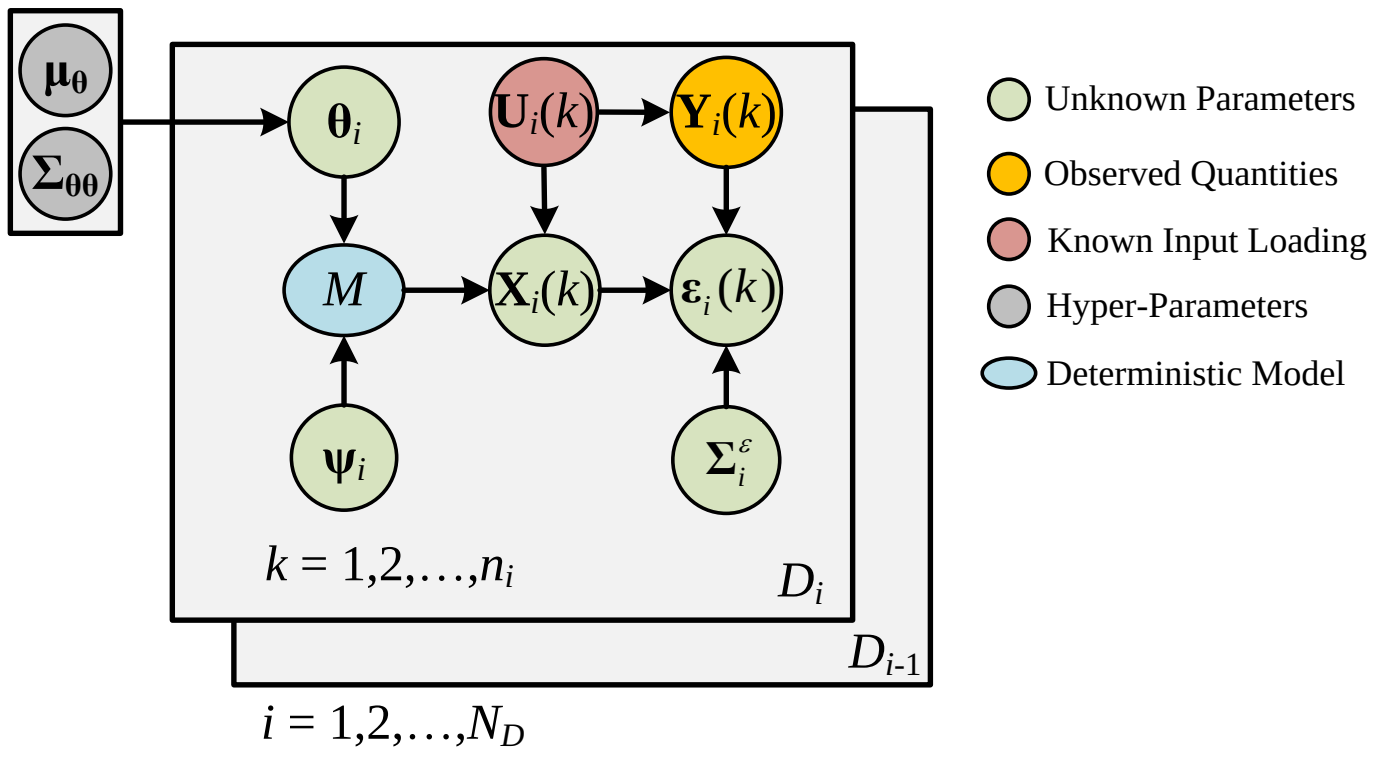

**Fig. 1.** Proposed multilevel probabilistic model

## 3. Uncertainty quantification

While the proposed multilevel model involves a number of parameters, the primary interest lies in updating the hyper-parameters. Therefore, the remaining parameters, including the prediction error parameters ($\mathbf{\Sigma}_i^{\varepsilon}$), the initial condition parameters ($\boldsymbol{\psi}_i$), and the model parameters ($\boldsymbol{\theta}_i$) are to be integrated out from the joint distribution as nuisance parameters. We begin this broad marginalization with the prediction error parameters that requires computing the following integration:

$$p\left(D_i \mid \boldsymbol{\theta}_i, \boldsymbol{\psi}_i\right) = \int_{\mathbf{\Sigma}_i^{\varepsilon}} p\left(D_i \mid \boldsymbol{\theta}_i, \boldsymbol{\psi}_i, \mathbf{\Sigma}_i^{\varepsilon}\right) p\left(\mathbf{\Sigma}_i^{\varepsilon}\right) d\mathbf{\Sigma}_i^{\varepsilon} \tag{12}$$

When $p\left(D_i \mid \boldsymbol{\theta}_i, \boldsymbol{\psi}_i, \mathbf{\Sigma}_i^{\varepsilon}\right)$ and $p\left(\mathbf{\Sigma}_i^{\varepsilon}\right)$ are respectively substituted from Eqs. (6) and (9), this integral will have the following explicit solution [12,34]:

$$p\left(D_i \mid \boldsymbol{\theta}_i, \boldsymbol{\psi}_i\right) \propto \exp\left(-L(\boldsymbol{\theta}_i, \boldsymbol{\psi}_i)\right) \tag{13}$$

and

$$L(\boldsymbol{\theta}_i, \boldsymbol{\psi}_i) = \frac{n_i}{2} \sum_{j=1}^{N_0} \ln\left[ \sum_{k=1}^{n_i} \left( \varepsilon_{i,j}(k\Delta t_i; \boldsymbol{\theta}_i, \boldsymbol{\psi}_i) \right)^2 \right] \tag{14}$$

where $L(\boldsymbol{\theta}_i, \boldsymbol{\psi}_i)$ is the negative logarithm of the marginalized likelihood function appearing as the summation over the logarithm of the prediction errors squares corresponding to each observed DOF. The probability distribution in Eq. (13) indicates the relative plausibility of different choices of $\boldsymbol{\theta}_i$ and $\boldsymbol{\psi}_i$. Those choices of $\boldsymbol{\theta}_i$ and $\boldsymbol{\psi}_i$ with smaller $L(\boldsymbol{\theta}_i, \boldsymbol{\psi}_i)$ correspond to greater likelihood, and those minimizing $L(\boldsymbol{\theta}_i, \boldsymbol{\psi}_i)$ (or maximizing the likelihood) are regarded as the most probable values with respect to each data set. When $L(\boldsymbol{\theta}_i, \boldsymbol{\psi}_i)$ is concentrated around one isolated peak, in the presence of a large number of data points the probability distribution in Eq. (13) can efficiently be approximated using a Laplace asymptotic approximation described as [11,12,35]

$$p\left(D_i \mid \boldsymbol{\theta}_i, \boldsymbol{\psi}_i\right) \approx N\left(\begin{bmatrix} \boldsymbol{\theta}_i \\ \boldsymbol{\psi}_i \end{bmatrix} \middle| \begin{bmatrix} \hat{\boldsymbol{\theta}}_i \\ \hat{\boldsymbol{\psi}}_i \end{bmatrix}, \begin{bmatrix} \hat{\mathbf{H}}_{\boldsymbol{\theta}_i\boldsymbol{\theta}_i} & \hat{\mathbf{H}}_{\boldsymbol{\theta}_i\boldsymbol{\psi}_i} \\ \hat{\mathbf{H}}^T_{\boldsymbol{\theta}_i\boldsymbol{\psi}_i} & \hat{\mathbf{H}}_{\boldsymbol{\psi}_i\boldsymbol{\psi}_i} \end{bmatrix}^{-1}\right) \tag{15}$$

and

$$\{\hat{\boldsymbol{\theta}}_i, \hat{\boldsymbol{\psi}}_i\} = \underset{\{\boldsymbol{\theta}_i, \boldsymbol{\psi}_i\}}{\operatorname{Argmin}} L(\boldsymbol{\theta}_i, \boldsymbol{\psi}_i) \tag{16}$$

$$\hat{\mathbf{H}}_{\boldsymbol{\theta}_i\boldsymbol{\theta}_i} = \nabla_{\boldsymbol{\theta}_i} \nabla^T_{\boldsymbol{\theta}_i} L(\hat{\boldsymbol{\theta}}_i, \hat{\boldsymbol{\psi}}_i), \;\; \hat{\mathbf{H}}_{\boldsymbol{\psi}_i\boldsymbol{\psi}_i} = \nabla_{\boldsymbol{\psi}_i} \nabla^T_{\boldsymbol{\psi}_i} L(\hat{\boldsymbol{\theta}}_i, \hat{\boldsymbol{\psi}}_i), \;\; \hat{\mathbf{H}}_{\boldsymbol{\theta}_i\boldsymbol{\psi}_i} = \nabla_{\boldsymbol{\theta}_i} \nabla^T_{\boldsymbol{\psi}_i} L(\hat{\boldsymbol{\theta}}_i, \hat{\boldsymbol{\psi}}_i) \tag{17}$$

Here, $\nabla(.)$ is the gradient operator; $\hat{\boldsymbol{\theta}}_i$ and $\hat{\boldsymbol{\psi}}_i$ are the MAP estimations; $\hat{\mathbf{H}}_{\boldsymbol{\theta}_i\boldsymbol{\theta}_i}$, $\hat{\mathbf{H}}_{\boldsymbol{\psi}_i\boldsymbol{\psi}_i}$, and $\hat{\mathbf{H}}_{\boldsymbol{\theta}_i\boldsymbol{\psi}_i}$ are the components of the Hessian matrix evaluated at the MAP estimations. The covariance matrix of the Gaussian distribution in Eq. (15) can be simplified using the well-known block matrix inversion lemma [36]. This optimization problem is known to yield reliable results when they are accompanied by analytical derivatives of the objective function with respect to the underlying parameters. In [12], the derivatives of $L(\boldsymbol{\theta}_i, \boldsymbol{\psi}_i)$ with respect to the parameters are obtained. Accordingly, computing such derivatives involve differentiating the model output quantities with respect to the model parameters. Numerous studies [15,37,38] have been dedicated to resolve this problem, obtaining analytical derivatives of model responses for both linear and nonlinear dynamical models.

Once the Gaussian approximation is obtained, we can marginalize the initial condition parameters in an explicit manner. Considering the prior distribution $p\left(\boldsymbol{\psi}_i\right)$ to be uniform, marginalizing $\boldsymbol{\psi}_i$ will only involve performing the integration over the Gaussian distribution in Eq. (15). For a set of parameters jointly described by a multivariate Gaussian distribution, the marginal distribution of each subset of the parameters also turns out to be Gaussian, where its mean and covariance matrix are determined by selecting the corresponding elements from the mean and covariance matrix of the joint distribution [39]. Accordingly, we can write:

$$p\left(D_i \mid \boldsymbol{\theta}_i\right) = \int_{\boldsymbol{\psi}_i} p\left(D_i \mid \boldsymbol{\theta}_i, \boldsymbol{\psi}_i\right) p\left(\boldsymbol{\psi}_i\right) d\boldsymbol{\psi}_i = N\left(\boldsymbol{\theta}_i \middle| \hat{\boldsymbol{\theta}}_i, \hat{\boldsymbol{\Sigma}}_{\boldsymbol{\theta}_i\boldsymbol{\theta}_i}\right) \tag{18}$$

where $\hat{\boldsymbol{\Sigma}}_{\boldsymbol{\theta}_i\boldsymbol{\theta}_i}$ is the covariance matrix of the marginal distribution computed as

$$\hat{\boldsymbol{\Sigma}}_{\boldsymbol{\theta}_i\boldsymbol{\theta}_i} = \left(\hat{\mathbf{H}}_{\boldsymbol{\theta}_i\boldsymbol{\theta}_i} - \hat{\mathbf{H}}_{\boldsymbol{\theta}_i\boldsymbol{\psi}_i}\left(\hat{\mathbf{H}}_{\boldsymbol{\psi}_i\boldsymbol{\psi}_i}\right)^{-1}\hat{\mathbf{H}}_{\boldsymbol{\theta}_i\boldsymbol{\psi}_i}^T\right)^{-1} \tag{19}$$

Having marginalized both $\boldsymbol{\Sigma}_i^{\varepsilon}$'s and $\boldsymbol{\psi}_i$'s, the joint distribution of the remaining parameters can described as

$$p\left(\left\{\boldsymbol{\theta}_i\right\}_{i=1}^{N_D}, \boldsymbol{\mu}_{\boldsymbol{\theta}}, \boldsymbol{\Sigma}_{\boldsymbol{\theta\theta}} \mid \mathbf{D}\right) \propto p\left(\boldsymbol{\mu}_{\boldsymbol{\theta}}, \boldsymbol{\Sigma}_{\boldsymbol{\theta\theta}}\right)\prod_{i=1}^{N_D}\left[N\left(\boldsymbol{\theta}_i \middle| \hat{\boldsymbol{\theta}}_i, \hat{\boldsymbol{\Sigma}}_{\boldsymbol{\theta}_i\boldsymbol{\theta}_i}\right) N\left(\boldsymbol{\theta}_i \mid \boldsymbol{\mu}_{\boldsymbol{\theta}}, \boldsymbol{\Sigma}_{\boldsymbol{\theta\theta}}\right)\right] \tag{20}$$

Now, marginalizing $\boldsymbol{\theta}_i$'s from this distribution is straightforward, which leads to the following analytical formulation for the marginal distribution of the hyper-parameters [12]:

$$p\left(\boldsymbol{\mu}_{\boldsymbol{\theta}}, \boldsymbol{\Sigma}_{\boldsymbol{\theta\theta}} \mid \mathbf{D}\right) \propto p\left(\boldsymbol{\mu}_{\boldsymbol{\theta}}, \boldsymbol{\Sigma}_{\boldsymbol{\theta\theta}}\right)\prod_{i=1}^{N_D} N\left(\boldsymbol{\mu}_{\boldsymbol{\theta}} \middle| \hat{\boldsymbol{\theta}}_i, \boldsymbol{\Sigma}_{\boldsymbol{\theta\theta}} + \hat{\boldsymbol{\Sigma}}_{\boldsymbol{\theta}_i\boldsymbol{\theta}_i}\right) \tag{21}$$

To obtain the MAP estimations, one can minimize the negative logarithm of the marginal distribution instead of maximizing this marginal probability distribution. Thus, the objective function is written as

$$\begin{aligned} L\left(\boldsymbol{\mu}_{\boldsymbol{\theta}}, \boldsymbol{\Sigma}_{\boldsymbol{\theta\theta}}\right) &= -\log p\left(\boldsymbol{\mu}_{\boldsymbol{\theta}}, \boldsymbol{\Sigma}_{\boldsymbol{\theta\theta}} \mid \mathbf{D}\right) \\ &= \frac{1}{2}\sum_{i=1}^{N_D}\log\left|\boldsymbol{\Sigma}_{\boldsymbol{\theta\theta}} + \hat{\boldsymbol{\Sigma}}_{\boldsymbol{\theta}_i\boldsymbol{\theta}_i}\right| + \frac{1}{2}\sum_{i=1}^{N_D}(\boldsymbol{\mu}_{\boldsymbol{\theta}} - \hat{\boldsymbol{\theta}}_i)^T(\boldsymbol{\Sigma}_{\boldsymbol{\theta\theta}} + \hat{\boldsymbol{\Sigma}}_{\boldsymbol{\theta}_i\boldsymbol{\theta}_i})^{-1}(\boldsymbol{\mu}_{\boldsymbol{\theta}} - \hat{\boldsymbol{\theta}}_i) - \ln p\left(\boldsymbol{\mu}_{\boldsymbol{\theta}}, \boldsymbol{\Sigma}_{\boldsymbol{\theta\theta}}\right) + c \end{aligned} \tag{22}$$

where $c$ is a constant value. Considering the covariance matrix $\boldsymbol{\Sigma}_{\boldsymbol{\theta\theta}}$ to be symmetric, the optimization will involve $N_\theta(N_\theta + 3)/2$ parameters representing the elements of both $\boldsymbol{\mu}_{\boldsymbol{\theta}}$ and $\boldsymbol{\Sigma}_{\boldsymbol{\theta\theta}}$. Moreover, the optimization can be performed reliably when analytical gradient of $L\left(\boldsymbol{\mu}_{\boldsymbol{\theta}}, \boldsymbol{\Sigma}_{\boldsymbol{\theta\theta}}\right)$ with respect to the hyper-parameters is supplied into the optimization toolbox. When $p\left(\boldsymbol{\mu}_{\boldsymbol{\theta}}, \boldsymbol{\Sigma}_{\boldsymbol{\theta\theta}}\right)$ is considered to be uniform, the first derivatives of $L\left(\boldsymbol{\mu}_{\boldsymbol{\theta}}, \boldsymbol{\Sigma}_{\boldsymbol{\theta\theta}}\right)$ with respect to the underlying parameters can be computed from

$$\frac{\partial L(\boldsymbol{\mu}_{\boldsymbol{\theta}}, \boldsymbol{\Sigma}_{\boldsymbol{\theta\theta}})}{\partial \boldsymbol{\mu}_{\boldsymbol{\theta}}} = \sum_{i=1}^{N_D} (\boldsymbol{\Sigma}_{\boldsymbol{\theta\theta}} + \hat{\boldsymbol{\Sigma}}_{\boldsymbol{\theta}_i\boldsymbol{\theta}_i})^{-1} (\boldsymbol{\mu}_{\boldsymbol{\theta}} - \hat{\boldsymbol{\theta}}_i) \tag{23}$$

$$\frac{\partial L(\boldsymbol{\mu}_{\boldsymbol{\theta}}, \boldsymbol{\Sigma}_{\boldsymbol{\theta\theta}})}{\partial \boldsymbol{\Sigma}_{\boldsymbol{\theta\theta}}} = \frac{1}{2} \sum_{i=1}^{N_D} \left[ (\boldsymbol{\Sigma}_{\boldsymbol{\theta\theta}} + \hat{\boldsymbol{\Sigma}}_{\boldsymbol{\theta}_i\boldsymbol{\theta}_i})^{-1} - (\boldsymbol{\Sigma}_{\boldsymbol{\theta\theta}} + \hat{\boldsymbol{\Sigma}}_{\boldsymbol{\theta}_i\boldsymbol{\theta}_i})^{-1} (\boldsymbol{\mu}_{\boldsymbol{\theta}} - \hat{\boldsymbol{\theta}}_i)(\boldsymbol{\mu}_{\boldsymbol{\theta}} - \hat{\boldsymbol{\theta}}_i)^T (\boldsymbol{\Sigma}_{\boldsymbol{\theta\theta}} + \hat{\boldsymbol{\Sigma}}_{\boldsymbol{\theta}_i\boldsymbol{\theta}_i})^{-1} \right] \tag{24}$$

Note that the MAP estimation of $\boldsymbol{\mu}_{\boldsymbol{\theta}}$ can be calculated directly from Eq. (23) giving:

$$\hat{\boldsymbol{\mu}}_{\boldsymbol{\theta}} = \sum_{i=1}^{N_D} \boldsymbol{\Lambda}_{\boldsymbol{\theta}_i} \hat{\boldsymbol{\theta}}_i \quad ; \quad \boldsymbol{\Lambda}_{\boldsymbol{\theta}_i} = \left[ \sum_{i=1}^{N_D} (\boldsymbol{\Sigma}_{\boldsymbol{\theta\theta}} + \hat{\boldsymbol{\Sigma}}_{\boldsymbol{\theta}_i\boldsymbol{\theta}_i})^{-1} \right]^{-1} (\boldsymbol{\Sigma}_{\boldsymbol{\theta\theta}} + \hat{\boldsymbol{\Sigma}}_{\boldsymbol{\theta}_i\boldsymbol{\theta}_i})^{-1} \tag{25}$$

where $\hat{\boldsymbol{\mu}}_{\boldsymbol{\theta}}$ is the MAP estimation of $\boldsymbol{\mu}_{\boldsymbol{\theta}}$, and $\boldsymbol{\Lambda}_{\boldsymbol{\theta}_i}$ is weighting matrices. This analytical expression allows it to write the optimization only in terms of the unknown covariance $\boldsymbol{\Sigma}_{\boldsymbol{\theta\theta}}$. Thus, the optimization can be carried out in two steps, where the MAP estimation of $\boldsymbol{\Sigma}_{\boldsymbol{\theta\theta}}$ is first obtained using Eq. (24) and will be followed by using Eq. (25) to calculate $\hat{\boldsymbol{\mu}}_{\boldsymbol{\theta}}$. Contrary to [14] that assumes $\boldsymbol{\Sigma}_{\boldsymbol{\theta\theta}}$ to be diagonal, this study does not postulate any prior structure on the hyper distribution covariance matrix.

Convergence and stability of this optimization problem can be further enhanced when reasonable initial estimations are considered. In this respect, an efficient approximation of the hyper-parameters' MAP estimations can be achieved when we neglect the variation of the covariance matrices $\hat{\boldsymbol{\Sigma}}_{\boldsymbol{\theta}_i\boldsymbol{\theta}_i}$ 's over data sets. This condition could be satisfied when the data sets have the same number of data points and the prediction errors maintain the same statistics over all data sets. Therefore, assuming $\hat{\boldsymbol{\Sigma}}_{\boldsymbol{\theta}_i\boldsymbol{\theta}_i} = \hat{\boldsymbol{\Sigma}}_{\boldsymbol{\theta}}$ , $\forall i = 1, 2, ..., N_D$ will provide the following explicit solutions:

$$\bar{\boldsymbol{\mu}}_{\boldsymbol{\theta}} = \frac{1}{N_D} \sum_{i=1}^{N_D} \hat{\boldsymbol{\theta}}_i \tag{26}$$

$$\bar{\mathbf{\Sigma}}_{\boldsymbol{\theta\theta}} = \left[ \frac{1}{N_D} \sum_{i=1}^{N_D} (\bar{\boldsymbol{\mu}}_{\boldsymbol{\theta}} - \hat{\boldsymbol{\theta}}_i)(\bar{\boldsymbol{\mu}}_{\boldsymbol{\theta}} - \hat{\boldsymbol{\theta}}_i)^T \right] - \hat{\mathbf{\Sigma}}_{\mathbf{0}} \tag{27}$$

where $\bar{\boldsymbol{\mu}}_{\boldsymbol{\theta}}$ and $\bar{\mathbf{\Sigma}}_{\boldsymbol{\theta\theta}}$ are the approximate hyper-parameters' MAP estimations. However, it should be noticed that the covariance matrix $\mathbf{\Sigma}_{\boldsymbol{\theta\theta}}$ must remain positive-definite over the whole optimization process. This can be implemented by imposing positive eigenvalues for $\mathbf{\Sigma}_{\boldsymbol{\theta\theta}}$ using a nonlinear constraint function.

These approximate solutions also suggest simplistic interpretations about the proposed hierarchical model. As indicated in Eq. (26), the estimated mean $\bar{\boldsymbol{\mu}}_{\boldsymbol{\theta}}$ is the average of $\hat{\boldsymbol{\theta}}_i$'s. When the covariance matrices $\hat{\mathbf{\Sigma}}_{\boldsymbol{\theta}_i\boldsymbol{\theta}_i}$'s are unequal, the estimated mean will turn into a weighted average of $\hat{\boldsymbol{\theta}}_i$'s, as obtained in Eq. (25). Moreover, as demonstrated in Eq. (27), the estimated covariance matrix is determined by subtracting the sample covariance matrix $\frac{1}{N_D} \sum_{i=1}^{N_D} (\bar{\boldsymbol{\mu}}_{\boldsymbol{\theta}} - \hat{\boldsymbol{\theta}}_i)(\bar{\boldsymbol{\mu}}_{\boldsymbol{\theta}} - \hat{\boldsymbol{\theta}}_i)^T$ from the data-set-specific covariance matrix $\hat{\mathbf{\Sigma}}_{\mathbf{0}}$. While the former expression reflects the variability over different data sets, the latter quantifies the uncertainty based on the mismatch between the actual and model responses. This interesting relationship between the sample mean and covariance matrix and the hyper-parameters MAP estimations, demonstrated for the first time, addresses relevant concerns about Bayesian model inference methods when dealing with multiple data sets.

The proposed Bayesian formulations are outlined in Algorithm 1. As presented, the algorithm begins with optimization of the data-set-specific parameters and is followed by computing the initial estimations of the hyper-parameters. Eventually, the hyper-parameters' MAP estimations are computed by the minimization of $L(\boldsymbol{\mu}_{\boldsymbol{\theta}}, \mathbf{\Sigma}_{\boldsymbol{\theta\theta}})$ with respect to both $\boldsymbol{\mu}_{\boldsymbol{\theta}}$ and $\mathbf{\Sigma}_{\boldsymbol{\theta\theta}}$. In the next section, we demonstrate how the uncertainty involved with the hyper-parameters can be propagated for response predictions.

**Algorithm 1**

Proposed hierarchical Bayesian uncertainty quantification and propagation method

---

**Uncertainty quantification**

1. For each data set (segment) $D_i$ , where $i = \{1, 2, ..., N_D\}$
    1.1. Minimize $L(\boldsymbol{\theta}_i, \boldsymbol{\psi}_i)$ given by Eq. (14) with respect to both $\boldsymbol{\theta}_i$ and $\boldsymbol{\psi}_i$
    1.2. Compute the MAP estimations $\hat{\boldsymbol{\theta}}_i$ and $\hat{\boldsymbol{\psi}}_i$
    1.3. Compute the Hessian matrix at the MAP estimations, i.e., $\hat{\mathbf{H}}_{\boldsymbol{\theta}_i\boldsymbol{\theta}_i}$, $\hat{\mathbf{H}}_{\boldsymbol{\theta}_i\boldsymbol{\psi}_i}$, $\hat{\mathbf{H}}_{\boldsymbol{\psi}_i\boldsymbol{\psi}_i}$
    1.4. Compute the covariance matrix $\hat{\boldsymbol{\Sigma}}_{\boldsymbol{\theta}_i\boldsymbol{\theta}_i}$ using Eq. (19)
2. End for
3. Choose $\hat{\boldsymbol{\Sigma}}_{\mathbf{0}}$ as the covariance matrix of one of the data sets
4. Compute the initial estimations of the hyper-parameters ( $\bar{\boldsymbol{\mu}}_{\boldsymbol{\theta}}$ and $\bar{\boldsymbol{\Sigma}}_{\boldsymbol{\theta\theta}}$ ) using Eqs. (26-27)
5. Minimize $L\left(\boldsymbol{\mu}_{\boldsymbol{\theta}}, \boldsymbol{\Sigma}_{\boldsymbol{\theta\theta}}\right)$ using the analytical gradient vector given in Eqs. (23-24)
6. Compute the MAP estimations of the hyper-parameters ( $\hat{\boldsymbol{\mu}}_{\boldsymbol{\theta}}$ and $\hat{\boldsymbol{\Sigma}}_{\boldsymbol{\theta\theta}}$ )

---

**Uncertainty propagation**

1. Choose the parameters of the prediction errors covariance matrix ( $\alpha_0$ and $\beta_0$ )
2. Draw samples $\boldsymbol{\theta}_{N_D+1}^{(m)}$ from $N\left(\boldsymbol{\theta}_{N_D+1} \,|\, \hat{\boldsymbol{\mu}}_{\boldsymbol{\theta}}, \hat{\boldsymbol{\Sigma}}_{\boldsymbol{\theta\theta}}\right)$, where $m = \{1, ..., N_s\}$
3. Compute second-moment statistics of the response using Eqs. (36-37)

---

## 4. Uncertainty Propagation

Once the deterministic model $M(\Phi)$ is updated based on multiple data sets, we can propagate the uncertainty to make predictions of output QoI when the system is subjected to future input loadings. Let $D_{N_D+1} = \{\mathbf{W}_{N_D+1}(k\Delta t_{N_D+1}) \in \mathbb{R}^{N_{DOF}}, k = 1, ..., n_{N_D+1}\}$ be the unobserved time-history system response that is to be predicted while $N_D$ data sets are already employed to calibrate the model. This dynamical response corresponds to the known input loading $U_{N_D+1} = \{\mathbf{U}_{N_D+1}(k\Delta t_{N_D+1}) \in \mathbb{R}^{N_I}, k = 1, ..., n_{N_D+1}\}$ and initial conditions $\hat{\boldsymbol{\psi}}_{N_D+1}$. The model $M(\Phi)$ produces time-history response $X_{N_D+1} = \{\mathbf{X}_{N_D+1}(k\Delta t_{N_D+1}; \boldsymbol{\theta}_{N_D+1}, \hat{\boldsymbol{\psi}}_{N_D+1}) \in \mathbb{R}^{N_{DOF}}, k = 1, ..., n_{N_D+1}\}$ under the input loading $U_{N_D+1}$ and the initial conditions $\hat{\boldsymbol{\psi}}_{N_D+1}$. Therefore, the model predicts the output time-history response as

$$\mathbf{W}_{N_D+1}(k\Delta t_{N_D+1}) = \mathbf{X}_{N_D+1}(k\Delta t_{N_D+1};\boldsymbol{\theta}_{N_D+1},\hat{\boldsymbol{\psi}}_{N_D+1}) + \boldsymbol{\varepsilon}_{N_D+1}(k\Delta t_{N_D+1};\boldsymbol{\theta}_{N_D+1},\hat{\boldsymbol{\psi}}_{N_D+1}) \tag{28}$$

where $\boldsymbol{\varepsilon}_{N_D+1}(.)$ is the unobserved prediction errors. Assuming the prediction errors of different DOF to be *i.i.d.* and describing those using Gaussian distributions leads to

$$\begin{aligned} p\left(\mathbf{W}_{N_D+1}(k\Delta t_{N_D+1}) \mid \boldsymbol{\theta}_{N_D+1},\hat{\boldsymbol{\psi}}_{N_D+1},\boldsymbol{\Sigma}^{\varepsilon}_{N_D+1}\right) & \\ = N\left(\mathbf{W}_{N_D+1}(k\Delta t_{N_D+1}) \mid \mathbf{X}_{N_D+1}(k\Delta t_{N_D+1};\boldsymbol{\theta}_{N_D+1},\hat{\boldsymbol{\psi}}_{N_D+1}),\boldsymbol{\Sigma}^{\varepsilon}_{N_D+1}\right) & \\ = \prod_{j=1}^{N_{DOF}} N\left(w_{N_D+1,j}(k\Delta t_{N_D+1}) \middle| x_{N_D+1,j}(k\Delta t_{N_D+1};\boldsymbol{\theta}_{N_D+1},\hat{\boldsymbol{\psi}}_{N_D+1}),\left(\sigma_{\varepsilon}^{(N_D+1,j)}\right)^2\right) & \end{aligned} \tag{29}$$

where $\boldsymbol{\Sigma}^{\varepsilon}_{N_D+1} \in \mathbb{R}^{N_{DOF}\times N_{DOF}}$ denotes the prediction error covariance matrix having diagonal entries $\left(\sigma_{\varepsilon}^{(N_D+1,j)}\right)^2, \forall j = 1,\ldots,N_{DOF}$ and zero off-diagonal entries; $w_{N_D+1,j}(.)$ and $x_{N_D+1,j}(.)$ denote the system and model responses of the $j^{\text{th}}$ DOF, respectively. Propagating the uncertainty of $\boldsymbol{\theta}_{N_D+1}$ and $\boldsymbol{\Sigma}^{\varepsilon}_{N_D+1}$ allows it to predict unobserved output QoI based on the uncertainty quantified earlier. Thus, the posterior predictive distribution of each data point can be determined from

$$\begin{aligned} p\left(\mathbf{W}_{N_D+1}(k\Delta t_{N_D+1}) \mid \mathbf{D}\right) & \\ = \int_{\boldsymbol{\theta}_{N_D+1}}\int_{\boldsymbol{\Sigma}^{\varepsilon}_{N_D+1}} p\left(\mathbf{W}_{N_D+1}(k\Delta t_{N_D+1}) \mid \boldsymbol{\theta}_{N_D+1},\hat{\boldsymbol{\psi}}_{N_D+1},\boldsymbol{\Sigma}^{\varepsilon}_{N_D+1}\right) p\left(\boldsymbol{\theta}_{N_D+1},\boldsymbol{\Sigma}^{\varepsilon}_{N_D+1} \mid \mathbf{D}\right) d\boldsymbol{\Sigma}^{\varepsilon}_{N_D+1} d\boldsymbol{\theta}_{N_D+1} & \end{aligned} \tag{30}$$

where $p\left(\boldsymbol{\theta}_{N_D+1},\boldsymbol{\Sigma}^{\varepsilon}_{N_D+1} \mid \mathbf{D}\right)$ can be decomposed into $p\left(\boldsymbol{\theta}_{N_D+1} \mid \mathbf{D}\right) p\left(\boldsymbol{\Sigma}^{\varepsilon}_{N_D+1}\right)$ since the prediction error covariance matrix is specific to the data sets, $p\left(\boldsymbol{\Sigma}^{\varepsilon}_{N_D+1}\right)$ is prior distribution of the covariance matrix, and $p\left(\boldsymbol{\theta}_{N_D+1} \mid \mathbf{D}\right)$ is computed as

$$p\left(\boldsymbol{\theta}_{N_D+1} \mid \mathbf{D}\right) = \int_{\boldsymbol{\Sigma}_{\boldsymbol{\theta\theta}}}\int_{\boldsymbol{\mu}_{\boldsymbol{\theta}}} p\left(\boldsymbol{\theta}_{N_D+1} \mid \boldsymbol{\mu}_{\boldsymbol{\theta}},\boldsymbol{\Sigma}_{\boldsymbol{\theta\theta}}\right) p\left(\boldsymbol{\mu}_{\boldsymbol{\theta}},\boldsymbol{\Sigma}_{\boldsymbol{\theta\theta}} \mid \mathbf{D}\right) d\boldsymbol{\mu}_{\boldsymbol{\theta}} d\boldsymbol{\Sigma}_{\boldsymbol{\theta\theta}} \approx N\left(\boldsymbol{\theta}_{N_D+1} \mid \hat{\boldsymbol{\mu}}_{\boldsymbol{\theta}},\hat{\boldsymbol{\Sigma}}_{\boldsymbol{\theta\theta}}\right) \tag{31}$$

This approximation neglects the uncertainty of the hyper-parameters as it substitutes the hyper-parameters with their MAP estimations ($\hat{\boldsymbol{\mu}}_{\boldsymbol{\theta}}$ and $\hat{\boldsymbol{\Sigma}}_{\boldsymbol{\theta\theta}}$). Regarding the probability distribution $p\left(\boldsymbol{\Sigma}^{\varepsilon}_{N_D+1}\right)$, we note that using non-informative prior distributions as the one introduced in Eq. (9) is inapplicable for response predictions since it leads to improper probability distributions. In this paper, we use inverse-gamma distributions to describe the prior distributions of the prediction error variances. This choice allows achieving explicit formulations for the integrations over the prediction errors variances that appeared earlier in Eq. (30). Therefore, we describe the prior distribution $p\left(\mathbf{\Sigma}^{\varepsilon}_{N_D+1}\right)$ as

$$p\left(\mathbf{\Sigma}^{\varepsilon}_{N_D+1}\right)=\prod_{j=1}^{N_{DOF}} IG\left(\left(\sigma_{\varepsilon}^{(N_D+1,j)}\right)^2 \mid \alpha_0,\beta_0\right) \tag{32}$$

where $IG(z\mid\alpha,\beta)=\dfrac{\beta^{\alpha}}{\Gamma(\alpha)}z^{-\alpha-1}\exp(-\beta/z)$ denotes Inverse-Gamma probability density functions. This choice of prior distribution is known to be conjugate for the variance of Gaussian distributions such that integrating $\left(\sigma_{\varepsilon}^{(N_D+1,j)}\right)^2$ over the interval $(0,+\infty)$ leads to [19]

$$p\left(\mathbf{W}_{N_D+1}(k\Delta t_{N_D+1})\mid\boldsymbol{\theta}_{N_D+1},\hat{\boldsymbol{\psi}}_{N_D+1}\right)=\prod_{j=1}^{N_{DOF}} t_{2\alpha_0}\left(w_{N_D+1,j}(k\Delta t_{N_D+1})\middle| x_{N_D+1,j}(k\Delta t_{N_D+1};\boldsymbol{\theta}_{N_D+1},\hat{\boldsymbol{\psi}}_{N_D+1}),\frac{\beta_0}{\alpha_0}\right) \tag{33}$$

where $t_{\nu}\left(x\middle|\mu,\sigma^2\right)$ denotes student's $t$-distribution with mean $\mu$, variance $\sigma^2$, and degree-of-freedom $\nu$. Thus, Eq. (30) can be rewritten as

$$\begin{aligned}&p\left(\mathbf{W}_{N_D+1}(k\Delta t_{N_D+1})\mid\mathbf{D}\right)\\&=\int_{\boldsymbol{\theta}_{N_D+1}}\prod_{j=1}^{N_{DOF}} t_{2\alpha_0}\left(w_{N_D+1,j}(k\Delta t_{N_D+1})\middle| x_{N_D+1,j}(k\Delta t_{N_D+1};\boldsymbol{\theta}_{N_D+1},\hat{\boldsymbol{\psi}}_{N_D+1}),\frac{\beta_0}{\alpha_0}\right)N\left(\boldsymbol{\theta}_{N_D+1}\mid\hat{\boldsymbol{\mu}}_{\boldsymbol{\theta}},\hat{\mathbf{\Sigma}}_{\boldsymbol{\theta\theta}}\right)d\boldsymbol{\theta}_{N_D+1}\end{aligned} \tag{34}$$

Except for the case that the response $x_{N_D+1,j}(k\Delta t_{N_D+1};\boldsymbol{\theta}_{N_D+1},\hat{\boldsymbol{\psi}}_{N_D+1})$ is a linear function of $\boldsymbol{\theta}_{N_D+1}$, arriving at closed-form solutions for this integral is not possible. However, using a sampling technique to draw samples $\boldsymbol{\theta}^{(m)}_{N_D+1}, m=\{1,2,...,N_s\}$, from the Gaussian distribution $N\left(\boldsymbol{\theta}_{N_D+1}\mid\hat{\boldsymbol{\mu}}_{\boldsymbol{\theta}},\hat{\mathbf{\Sigma}}_{\boldsymbol{\theta\theta}}\right)$ will provide an approximate solution as follows:

$$p\left(\mathbf{W}_{N_D+1}(k\Delta t_{N_D+1})\mid\mathbf{D}\right)=\frac{1}{N_s}\sum_{m=1}^{N_s}\prod_{j=1}^{N_{DOF}} t_{2\alpha_0}\left(w_{N_D+1,j}(k\Delta t_{N_D+1})\middle| x_{N_D+1,j}(k\Delta t_{N_D+1};\boldsymbol{\theta}^{(m)}_{N_D+1},\hat{\boldsymbol{\psi}}_{N_D+1}),\frac{\beta_0}{\alpha_0}\right) \tag{35}$$

where $N_s$ is the number of samples. This posterior predictive distribution allows predicting system response QoI when the loading and initial conditions are given. An interesting feature of student's *t*-distributions is the gentle tails it produces [19] as compared to the Gaussian predictive distributions suggested in [1].

Sedehi et al. [12] have proved that the second-moment statistics of distributions expressed as the multiplication of student's *t*-distributions can be computed as

$$E\left(\mathbf{W}_{N_D+1}(k\Delta t_{N_D+1})\,|\,\mathbf{D}\right)=\frac{1}{N_s}\sum_{m=1}^{N_s}\mathbf{X}_{N_D+1}(k\Delta t_{N_D+1};\boldsymbol{\theta}_{N_D+1}^{(m)},\hat{\boldsymbol{\psi}}_{N_D+1}) \tag{36}$$

$$\begin{aligned}
&CoV\left(\mathbf{W}_{N_D+1}(k\Delta t_{N_D+1})\,|\,\mathbf{D}\right)\\
&=\frac{1}{N_s}\sum_{m=1}^{N_s}\left[\mathbf{X}_{N_D+1}(k\Delta t_{N_D+1};\boldsymbol{\theta}_{N_D+1}^{(m)},\hat{\boldsymbol{\psi}}_{N_D+1})\mathbf{X}_{N_D+1}^{T}(k\Delta t_{N_D+1};\boldsymbol{\theta}_{N_D+1}^{(m)},\hat{\boldsymbol{\psi}}_{N_D+1})-\frac{2\alpha_0}{2\alpha_0-2}\left(\frac{\beta_0}{\alpha_0}\right)\mathbf{I}_{N_{DOF}}\right]\\
&-\left(\frac{1}{N_s}\sum_{m=1}^{N_s}\mathbf{X}_{N_D+1}(k\Delta t_{N_D+1};\boldsymbol{\theta}_{N_D+1}^{(m)},\hat{\boldsymbol{\psi}}_{N_D+1})\right)\left(\frac{1}{N_s}\sum_{m=1}^{N_s}\mathbf{X}_{N_D+1}(k\Delta t_{N_D+1};\boldsymbol{\theta}_{N_D+1}^{(m)},\hat{\boldsymbol{\psi}}_{N_D+1})\right)^{T}
\end{aligned} \tag{37}$$

where $\mathbf{I}_{N_{DOF}}\in\mathbb{R}^{N_{DOF}\times N_{DOF}}$ is identity matrix; $E(.)$ and $CoV(.)$ denote the mean and covariance matrix, respectively. The procedure offered for predicting response QoI is summarized in Algorithm 1.

## 5. Illustrative Examples

### 5.1. Single-degree of freedom (SDOF) system

A SDOF dynamical system is chosen to create synthetic time-history response. The system is considered to have the natural frequency $\overline{f}$ and 5% viscous damping ratio. Due to unknown nonlinear mechanisms, the natural frequency is assumed to follow $N\left(\overline{f}\,|\,(2\pi)^{-1},(200\pi)^{-2}\right)$. The input force is Gaussian white noise (GWN) base excitation shown in Fig. 2(a). The sampling rate and the spectral power of the input excitation are 200Hz and 0.0013m$^2$/s$^4$, respectively. The initial conditions, including the initial velocity and displacement, are both zero. The time-history displacement response is 2000s long sampled at $\Delta t = 0.005s$ intervals. To account for the measurement noise, additive white Gaussian noise (AWGN) is introduced such that the root mean square (RMS) of the noise will be 1% of the RMS of the noise-free response.

The structural model used to describe the system is a linear SDOF system with an unknown natural frequency ( $f$ ) that should be inferred from the data. The viscous damping ratio ($\xi$) is considered to be 4.5% that aims at imposing modeling errors as the actual damping is 5%. The natural frequency is the only uncertain parameter $\boldsymbol{\theta} \equiv f$ modeled using the Gaussian distribution $N(f \mid \mu_f, \sigma_f^2)$, where $\phi = \{\mu_f, \sigma_f\}$ is the set comprising uncertain hyper-parameters. The prior distribution $p(\phi)$ is assumed to be uniform described as $\mu_f \sim U(0,3)$ and $\sigma_f^2 \sim U(0,1)$. The input force is considered to be known. However, the initial displacement and velocity are unknown parameters subsumed into $\boldsymbol{\psi} \equiv [y_0 \quad \dot{y}_0]^T$. Hence, the structural model can be characterized as $M(f, y_0, \dot{y}_0)$, which should be calibrated using multiple segments of noisy time-history displacement response. For this purpose, the system response is broken into $N_D = 40$ segments each 50s long. Let $\theta = \{\boldsymbol{\theta}_i, i = 1,..., N_D\}$ and $\psi = \{\boldsymbol{\psi}_i, i = 1,..., N_D\}$ comprise the segment-specific parameters. The prior distribution $p(\boldsymbol{\psi}_i)$ is described using the uniform distribution $y_{0,i}, \dot{y}_{0,i} \sim U(-20, 20)$. It should be noticed that the segments length is chosen sufficiently long so that the global identifiability of $\boldsymbol{\theta}_i$ 's and $\boldsymbol{\psi}_i$ 's can be assured.

The asymptotic approximation obtained in Eqs. (13-15) is used for inferring $\boldsymbol{\theta}_i$ 's and $\boldsymbol{\psi}_i$ 's from each segment of data. Fig. 2(b) shows the model response created using the MAP estimations of $\boldsymbol{\theta}_i$ 's and $\boldsymbol{\psi}_i$ 's along with the actual noisy response. The two curves are in close agreement demonstrating the validity of the model inference. Once the MAP estimations and the associated uncertainty are computed for each segment, one can compute initial estimations of the hyper-parameters using the procedure suggested through Eqs. (26-27). Thus, we obtain $\bar{\mu}_f = 0.1595\text{Hz}$ and $\bar{\sigma}_f = 0.0017\text{Hz}$ as the initial estimations. Using them for optimizing the objective function in Eq. (22) with respect to the hyper-parameters provides the MAP estimations $\hat{\mu}_f = 0.1595\text{Hz}$ and $\hat{\sigma}_f = 0.00169\text{Hz}$. As demonstrated, the initial estimations are very close to the solution of the optimization problem such that one can accept them as the optimal values. These MAP estimations are also evident on $p(\mu_f, \sigma_f \mid \mathbf{D})$ plot depicted in Fig. 3(a).

The Gaussian distributions $N\left(f_i \mid \hat{f}_i, \hat{\sigma}^2_{f_i}\right)$, where $i = \{1, 2, ..., N_D\}$, obtained from each segment of data are plotted in Fig. 3(b) using the blue curves. As indicated, the segment-specific realizations of the frequency draw sharp peaks at different frequencies. Nevertheless, the associated uncertainties are extremely small and do not account for the variability of $\hat{f}_i$'s over different segments. The hyper distribution $N(f \mid \hat{\mu}_f, \hat{\sigma}^2_f)$ is also shown in this figure using the red curve. Contrary to the segment-specific realizations, the proposed hierarchical approach provides robust uncertainty and captures the variability realistically.

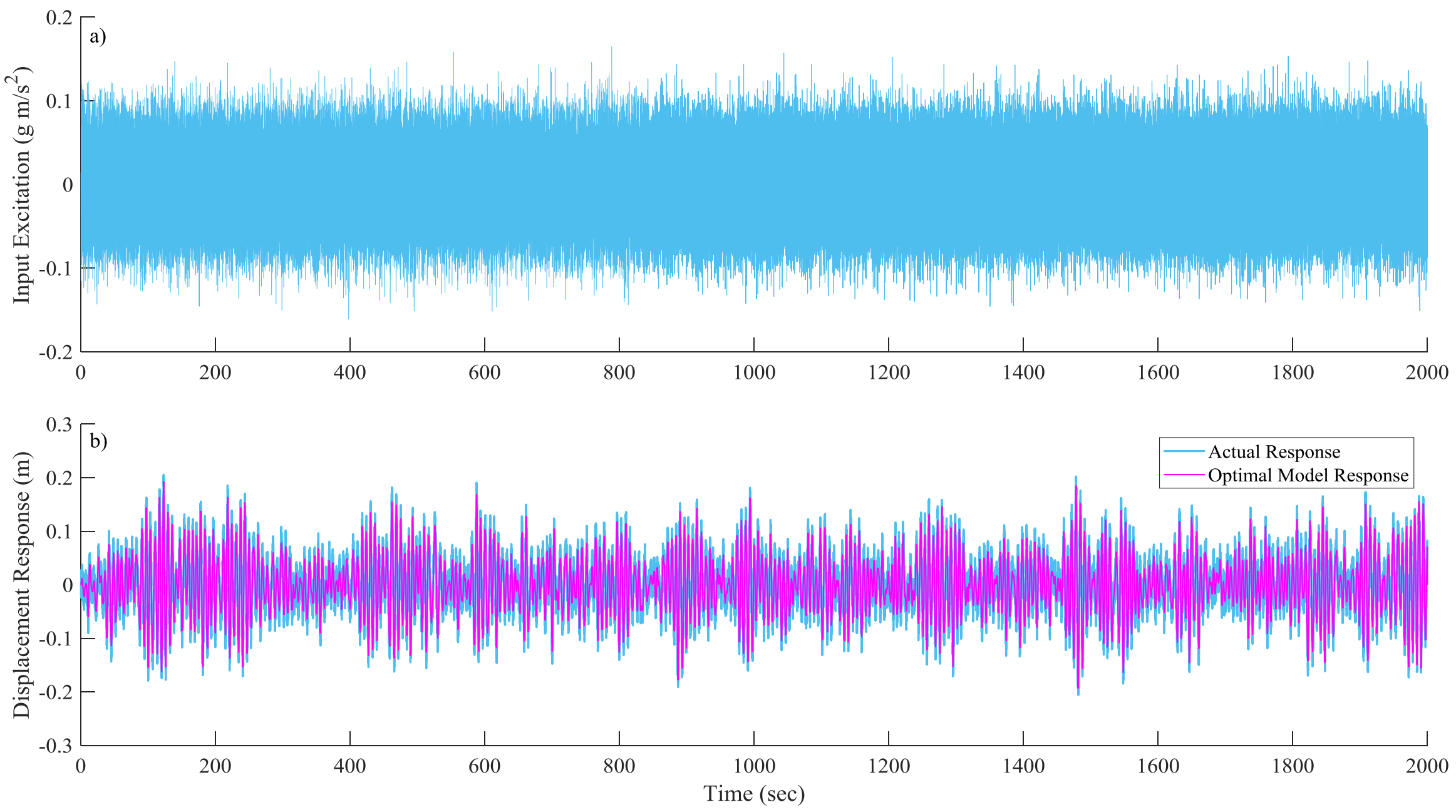

**Fig. 2.** (a) GWN input excitation (b) Noisy displacement response of the SDOF system compared with the response of an optimal model ($f$ = 0.1595Hz) calibrated using the proposed second-order approximation

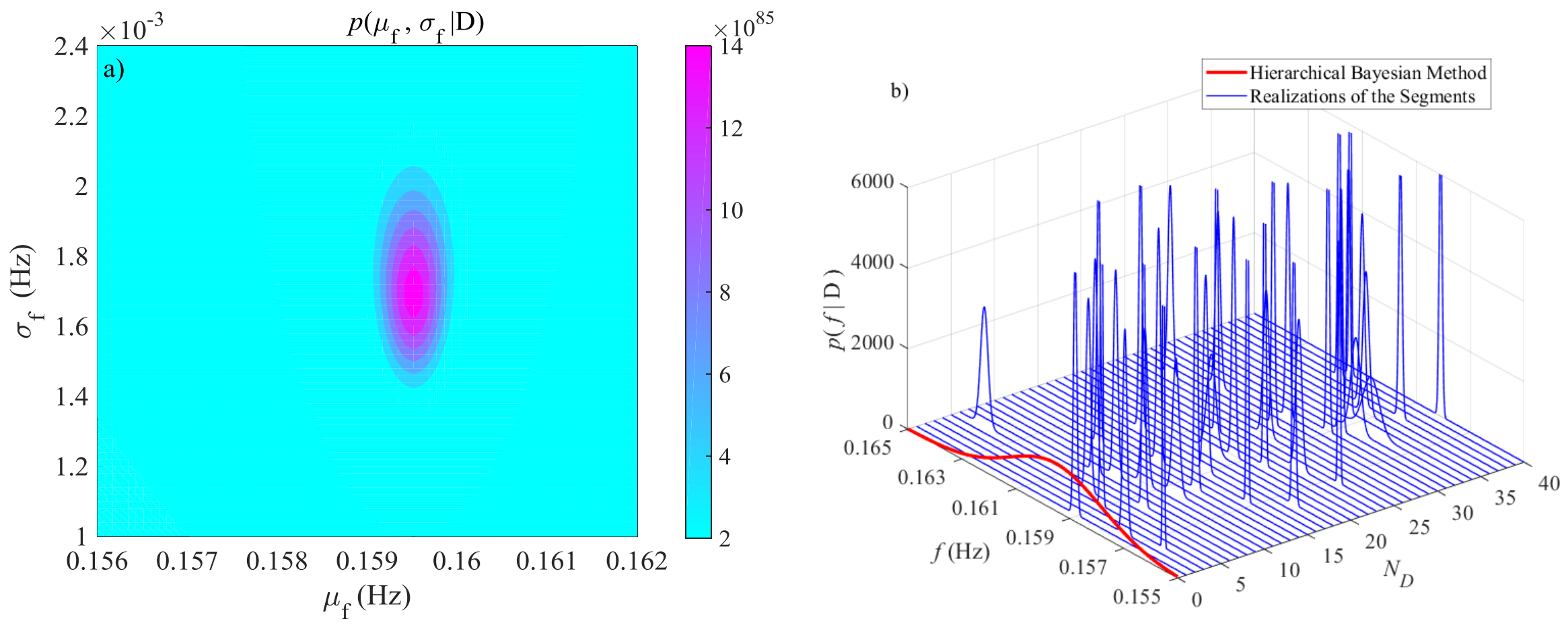

**Fig. 3.** (a) Posterior distribution of the hyper-parameters (b) comparison between the Gaussian distributions obtained from each segment of data indicated by blue curves and the hyper probability distribution indicated by the red curve

Another GWN input excitation having the same properties as the excitation used in the model calibration phase is applied to the SDOF system. The length of this input is considered to be 50s sampled at 0.005s intervals. The system response under this input is to be predicted using the uncertainty propagation method outlined in Algorithm 1. For this purpose, the prediction error parameters are set to $\alpha_0 = 2$ and $\beta_0 \to 0$. This limiting choice suggested in [19] allows it to neglect the contribution of prediction error parameters for response predictions. Subsequently, the samples $f_{N_D+1}^{(m)}$, $m = \{1, \ldots, N_s = 2000\}$, are drawn from the hyper distribution $N\left( f_{N_D+1} | \hat{\mu}_f, \hat{\sigma}_f^2 \right)$. Using Eqs. (36-37) yields the second-moment statistics of the predicted responses. Fig. 4 shows predictions of displacement, velocity, and acceleration time-history responses along with the actual responses. The predicted mean response is in good agreement with the actual response. The uncertainty bound associated with the displacement and velocity responses grow over time accounting for error accumulation effects. At the same time, when the acceleration response is predicted, the uncertainty bound remains reasonably small due to superb accuracy in predictions.

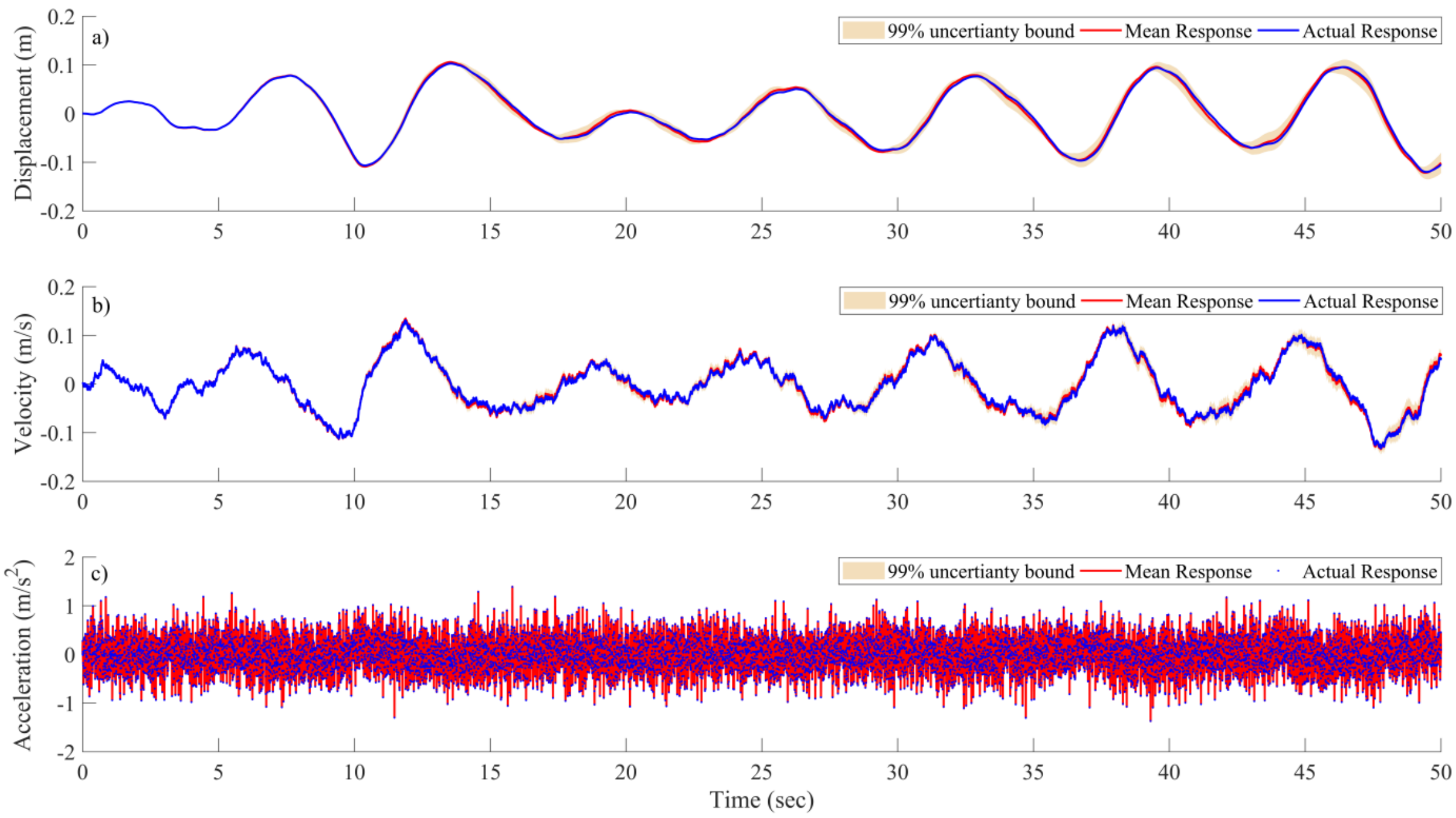

**Fig. 4.** Predictions of time-history displacement, velocity, and acceleration responses

The results presented earlier are obtained by breaking the 2000s time-history data into 40 segments, each 50s long. Table 1 presents the MAP estimations of the hyper-parameters when different groupings are considered. According to this table, the estimated uncertainties slightly change when having different choices of the number of segments and data points, provided a sufficiently large number of segments and data points. Therefore, the presented results will remain almost the same regardless of the choice of the number of segments and data points, and the groupings have minimal impact on the computed uncertainties.

**Table 1.**
MAP estimations of the hyper-parameters for different choices of the number of data points and segments

| $n_i$ [1] | $L_i$ [2](s) | $N_D$=20 | | $N_D$=40 | | $N_D$=50 | |
|---|---|---|---|---|---|---|---|
| | | $\hat{\mu}_f$ | $\hat{\sigma}_f$ | $\hat{\mu}_f$ | $\hat{\sigma}_f$ | $\hat{\mu}_f$ | $\hat{\sigma}_f$ |
| 1000 | 5 | 0.1593 | 0.0009 | 0.1592 | 0.0010 | 0.1592 | 0.0011 |
| 2000 | 10 | 0.1594 | 0.0014 | 0.1592 | 0.0014 | 0.1591 | 0.0015 |
| 4000 | 20 | 0.1594 | 0.0015 | 0.1594 | 0.0014 | 0.1593 | 0.0015 |
| 8000 | 40 | 0.1597 | 0.0018 | 0.1599 | 0.0016 | 0.1598 | 0.0016 |

[1] number of data point within each segment

[2] length of each segment

## 5.2. Experimental example

### *5.2.1. Modeling assumptions*

A three-story shear building prototype structure is tested on a shaking-table under GWN base excitation to test and verify the proposed method. The prototype is shown in Fig. 5(a). We use incomplete input-output vibrational measurements of the prototype to demonstrate the proposed hierarchical method. Although full acceleration time-history responses are measured, we only use those corresponding to the base and the third floor when the structure undergoes GWN base excitation. The measured response is divided into $N_D$=98 segments, each 10s long and sampled at 200Hz rate. Full details of the experimental setup, dynamical characteristics of the prototype, and modeling assumptions can be found elsewhere [12,40]. For completeness, however, we quickly review the following characteristics and assumptions:

- The mass of the first, second, and third floors are $m_1$=5.63kg, $m_2$=6.03kg, $m_3$=4.66kg, respectively. These values are obtained by direct measurements and reported in [40].
- The nominal stiffness of each floor is reported as $k_1$=20.88kN/m, $k_2$=22.37kN/m, and $k_3$=24.21kN/m, respectively [40]. Note that these nominal values correspond to the stiffness of the equivalent shear building model shown in Fig. 5(b).

- The mode frequencies and damping ratios of the three dynamical modes are computed as ($f_1$=4.23Hz, $\xi_1$=2.39%), ($f_2$=12.78Hz, $\xi_2$=0.87%), and ($f_3$=18.65Hz, $\xi_3$=0.65%), respectively [40]. These modal properties are estimated by applying a spectral density approach and using a classical damping model described in [40].

The dynamical model is a linear shear building model having 3 DOF shown in Fig. 5(b). The mass matrix is assumed to be known and diagonal. The stiffness matrix is expressed using shear frame assumptions giving:

$$\mathbf{K} = \begin{bmatrix} k_1\theta_1 + k_2\theta_2 & -k_2\theta_2 & 0 \\ -k_2\theta_2 & k_2\theta_2 + k_3\theta_3 & -k_3\theta_3 \\ 0 & -k_3\theta_3 & k_3\theta_3 \end{bmatrix} \tag{38}$$

Using the damping model suggested in [41] gives:

$$\mathbf{C} = \left( 4\pi f_1\xi_1 \frac{\mathbf{M}\phi_1\phi_1^T\mathbf{M}}{\phi_1^T\mathbf{M}\phi_1} \right)\theta_4 + \left( 4\pi f_2\xi_2 \frac{\mathbf{M}\phi_2\phi_2^T\mathbf{M}}{\phi_2^T\mathbf{M}\phi_2} \right)\theta_5 + \left( 4\pi f_3\xi_3 \frac{\mathbf{M}\phi_3\phi_3^T\mathbf{M}}{\phi_3^T\mathbf{M}\phi_3} \right)\theta_6 \tag{39}$$

where $\mathbf{K} \in \mathbb{R}^{3\times3}$ and $\mathbf{C} \in \mathbb{R}^{3\times3}$ are the stiffness and damping matrices, respectively; $\mathbf{M} \in \mathbb{R}^{3\times3}$ is the diagonal mass matrix having diagonal entries $m_1$, $m_2$, and $m_3$ described earlier; $\boldsymbol{\theta} = [\theta_1 \;\; \theta_2 \;\; \theta_3 \;\; \theta_4 \;\; \theta_5 \;\; \theta_6]^T$ comprises the unknown parameters; $f_i$, $\xi_i$, and $\phi_i$ are the nominal mode frequencies, modal damping ratios, and mode shapes, respectively. Calibration of the parameter vector $\boldsymbol{\theta}$ from multiple data segments is the primary goal of this study. However, having unknown initial conditions will require introducing and calibrating the parameter vector $\boldsymbol{\psi} = [\psi_1 \;\; \psi_2 \;\; \psi_3 \;\; \psi_4 \;\; \psi_5 \;\; \psi_6]^T$ for each segment of data, where $\psi_1$, $\psi_2$, and $\psi_3$ are the initial displacements and $\psi_4$, $\psi_5$, and $\psi_6$ are the initial velocities corresponding to the 1st, 2nd, and 3rd DOF, respectively. Note that both $\boldsymbol{\theta}$ and $\boldsymbol{\psi}$ are considered to be variable over data segments. Besides, the variation of $\boldsymbol{\theta}$ over multiple data segments is demonstrated using the subscript $i$, described by the hyper-prior probability distribution $N\left(\boldsymbol{\theta}_i \mid \boldsymbol{\mu}_{\boldsymbol{\theta}}, \boldsymbol{\Sigma}_{\boldsymbol{\theta\theta}}\right)$, where $\boldsymbol{\mu}_{\boldsymbol{\theta}} \in \mathbb{R}^{6\times1}$ and $\boldsymbol{\Sigma}_{\boldsymbol{\theta\theta}} \in \mathbb{R}^{6\times6}$ are the mean and covariance matrix that should be updated from the data.

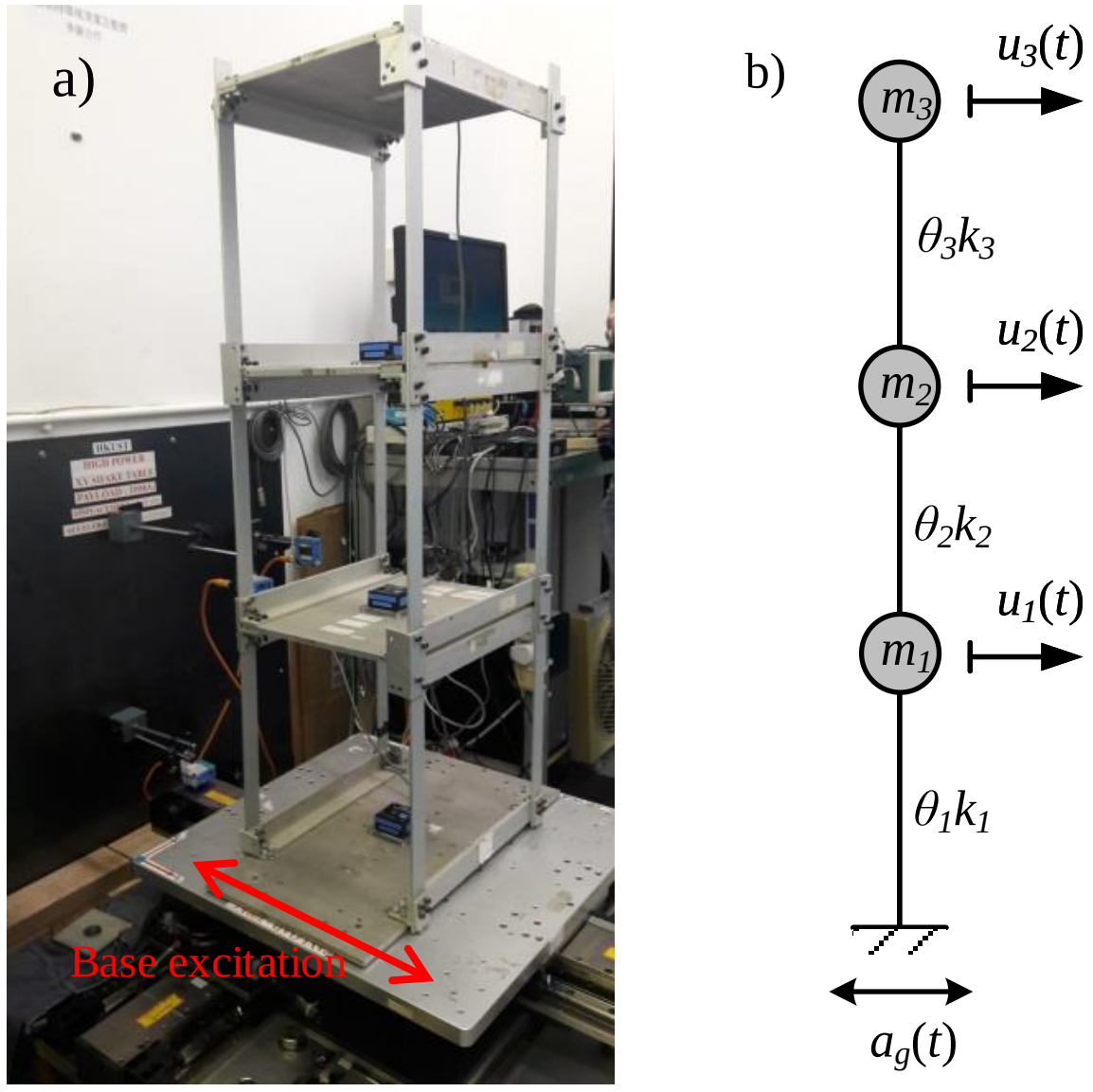

**Fig. 5.** (a) Three-story structure prototype tested on a shaking table (b) shear-building dynamical model

*5.2.2. Model uncertainty*

Following the procedure summarized in Algorithm 1, the optimization was carried out particularly for each segment of data. Thus, the MAP estimations of both $\boldsymbol{\theta}_i$'s and $\boldsymbol{\psi}_i$'s along with their posterior uncertainty are obtained. Fig. 6 plots the first floor acceleration response obtained by substituting the MAP estimations into the model characteristic equations. This results is compared with the measured response of six arbitrary 10s time-history segments. The accuracy of model predictions is good, which verifies the modeling assumptions used for describing the vibrational response. Combining segment-specific realizations under the proposed hierarchical setting gives the marginal posterior distribution of the hyper-parameters. Performing the optimization formulated in Eqs. (22-25) provides the hyper-parameters' MAP estimations, as summarized in Table 2. The MAP estimation of the mean vector is presented under "mean" and indicated by $\hat{\mu}_{\theta_p}$. The diagonal entries of the covariance matrix $\boldsymbol{\Sigma}_{\boldsymbol{\theta\theta}}$ are indicated under $\hat{\sigma}^2_{\theta_p}$ column, and the off-diagonal entries are decomposed as $\hat{\sigma}^2_{\theta_p\theta_q} = \hat{\rho}_{\theta_p\theta_q}\hat{\sigma}_{\theta_p}\hat{\sigma}_{\theta_q}$, where $\hat{\rho}_{\theta_p\theta_q}$ represents the correlation between $\theta_p$ and $\theta_q$ that can also be found in Table 2. Accordingly, there is high negative correlation between the stiffness parameters $\theta_1$ and $\theta_2$, as well as $\theta_2$ and $\theta_3$. Having negative correlation makes intuitive sense as reducing one element increases the other one. However, the correlation between damping parameters, $\theta_4$, $\theta_5$, and $\theta_6$, is almost negligible. This can be due to the fact that they correspond to independent mode shapes. The correlation between the stiffness and damping parameters vary from moderate to negligible. It is also worth to note that the uncertainty involved with the damping parameters ($\theta_4, \theta_5, \theta_6$) is far greater that the uncertainty involved with the stiffness parameters ( $\theta_1, \theta_2, \theta_3$). This finding can be attributed to the inaccuracy of the damping model.

Fig. 7 demonstrates the hyper distribution $N(\boldsymbol{\theta} \mid \hat{\boldsymbol{\mu}}_{\boldsymbol{\theta}}, \hat{\boldsymbol{\Sigma}}_{\boldsymbol{\theta}})$ in a matrix-plot format. The upper triangular elements show the joint distribution of parameter pairs $p(\theta_p, \theta_q \mid \mathbf{D})$, where $p, q = \{1, ..., 6\}$. The plots on the diagonal indicate the marginal distributions $p(\theta_p \mid \mathbf{D})$, and the lower triangular figures show the MAP estimations of $\boldsymbol{\theta}_i$'s in a scatter layout. The marginal distributions on the diagonal give robust uncertainty with respect to the variability over different segments. It is also evident that the correlation pattern appearing in the scatter plots agrees very well with the corresponding from the upper triangular contour plots. These findings clearly indicate the dominant impact of the inherent variability on the calibrated hyper distribution.

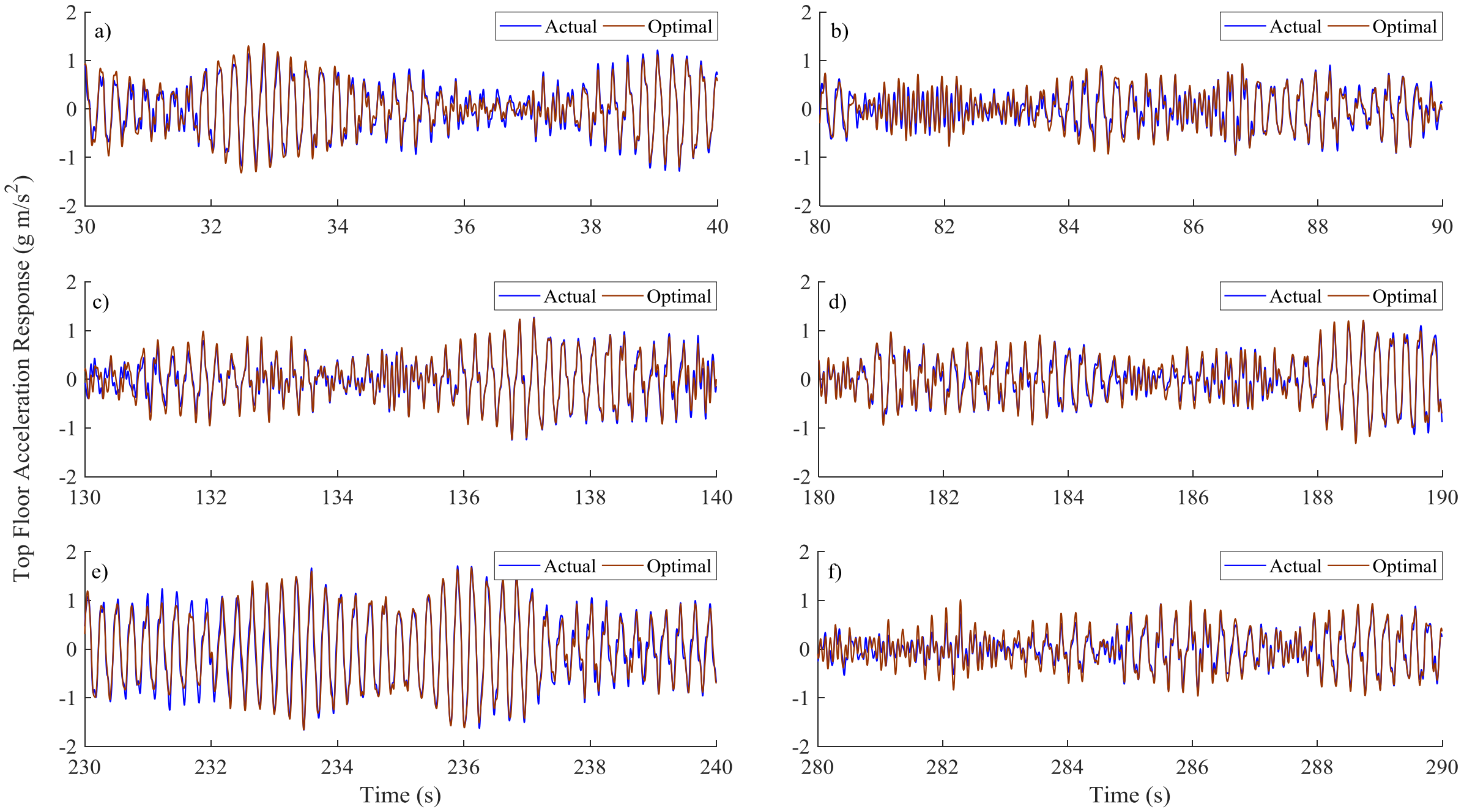

**Fig. 6.** Measured time-history acceleration response of the third floor along with the corresponding model predictions generated using the segment-specific optimal realizations (Six arbitrary segments are shown)

**Table 2.**
MAP estimation of the hyper-parameters computed using Algorithm 1

| | Mean | Variance | Correlation coefficients | | | | |
|---|---|---|---|---|---|---|---|
| $\theta_p$ | $\hat{\mu}_{\theta_p}$ | $\hat{\sigma}^2_{\theta_p}$ | $\hat{\rho}_{\theta_p\theta_2}$ | $\hat{\rho}_{\theta_p\theta_3}$ | $\hat{\rho}_{\theta_p\theta_4}$ | $\hat{\rho}_{\theta_p\theta_5}$ | $\hat{\rho}_{\theta_p\theta_6}$ |
| $\theta_1$ | 0.8274 | 0.0002 | -0.7894 | 0.5752 | 0.2740 | 0.3205 | 0.0026 |
| $\theta_2$ | 1.1055 | 0.0022 | | -0.9166 | -0.3360 | -0.3204 | 0.0109 |
| $\theta_3$ | 1.0766 | 0.0009 | | | 0.2160 | 0.3211 | -0.0240 |
| $\theta_4$ | 1.0745 | 0.4321 | | | | 0.0434 | -0.0323 |
| $\theta_5$ | 0.4242 | 0.0295 | | | | | 0.0727 |
| $\theta_6$ | 1.1265 | 0.1462 | | | | | |

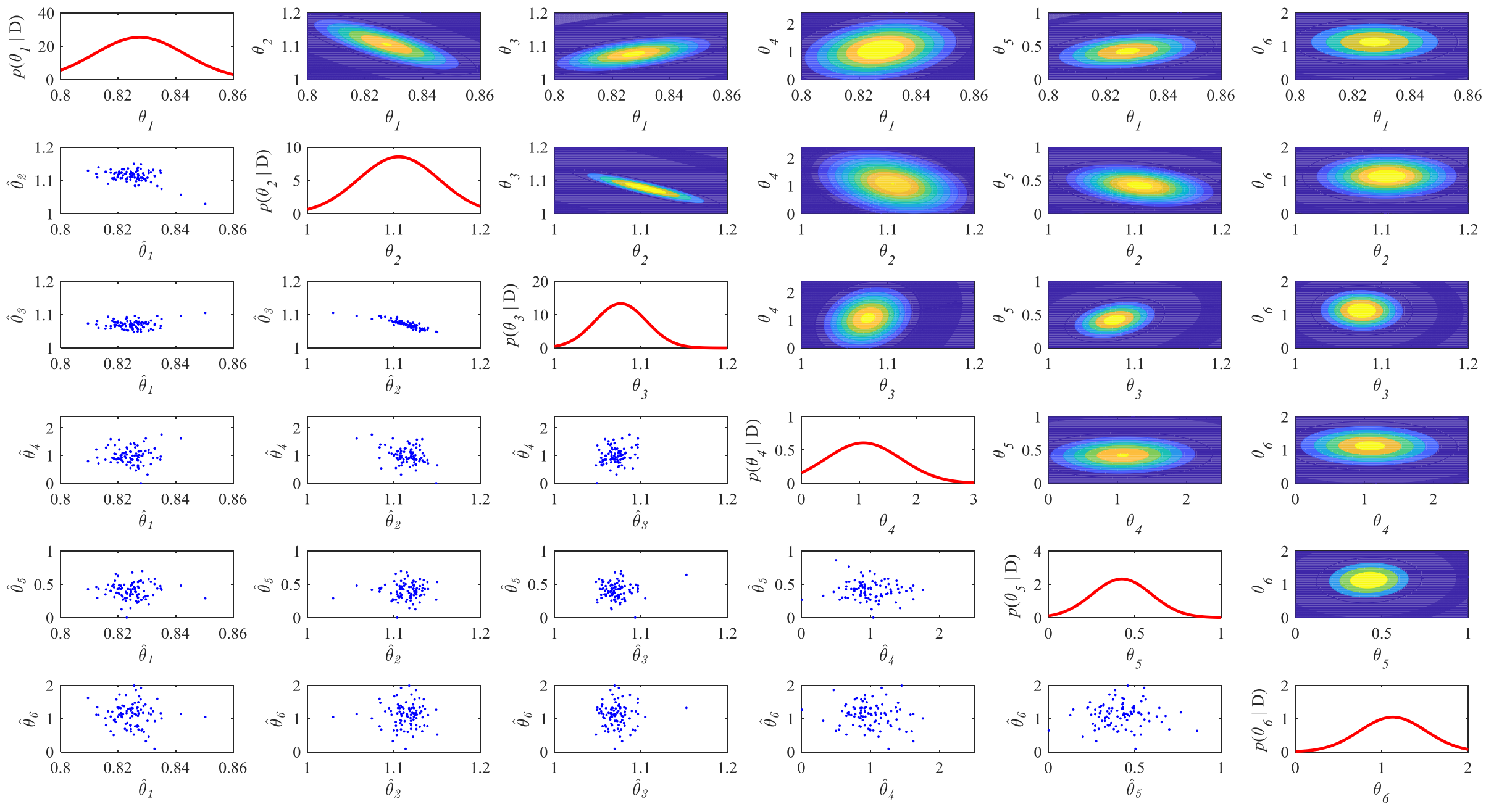

**Fig. 7.** Uncertainty quantification using the proposed hierarchical method (Plots on the diagonal show the marginal distributions $p(\theta_p \mid \mathbf{D})$; the upper triangular plots show contours of the joint distributions $p(\theta_p, \theta_q \mid \mathbf{D})$; the lower triangular plots show scatter plots of $\hat{\theta}_p$'s vs. $\hat{\theta}_q$'s)

### *5.2.3. Response predictions*

The proposed formulations are used to predict the displacement, velocity, and acceleration responses under a new GWN base excitation while the system is initially at rest. Alike the SDOF example, the limiting choice $\alpha_0 = 2$ and $\beta_0 \to 0$ is made for the prediction error parameters. Therefore, we propagate only the parametric uncertainty described by the hyper distribution $N(\boldsymbol{\theta} \mid \hat{\boldsymbol{\mu}}_{\boldsymbol{\theta}}, \hat{\boldsymbol{\Sigma}}_{\boldsymbol{\theta}})$. Drawing the random samples $\boldsymbol{\theta}^{(m)}$, $m = \{1, 2, ..., N_s = 2000\}$, from this multivariate Gaussian distribution can be accomplished using MATLAB library functions [42]. The mean and covariance matrix of the response can thus be computed using Eqs. (36-37). Fig. 8 shows the mean of the predicted response corresponding to the first floor along with 99% uncertainty bounds. The mean and the actual responses are in good agreement, and the uncertainty bound appears to be robust as it completely captures the discrepancy between the mean and actual response.

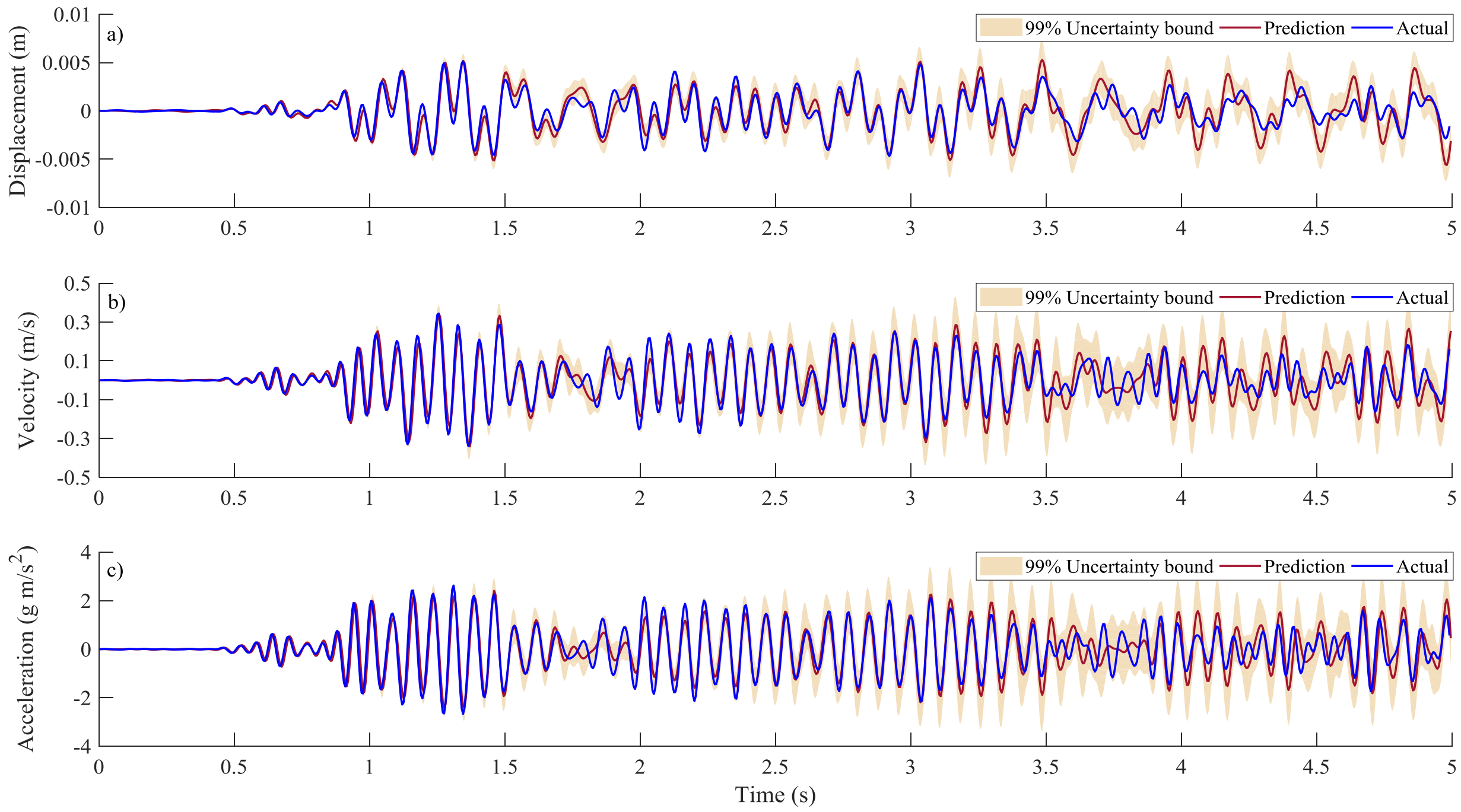

**Fig. 8.** Predictions of the first floor time-history responses

## 6. Conclusions

Data-driven uncertainty quantification and propagation of dynamical models is addressed through developing a novel hierarchical Bayesian framework. While the state-of-the-art Bayesian methods often use stationary assumptions for the parameters, the proposed framework captures it through breaking time-history vibrational data into non-overlapping segments. The segments and parameters thereof are assumed to be statistically independent. The segment-specific parameters include the model's dynamical parameters, as well as the initial conditions and prediction error parameters. While the latter is required to be marginalized out, the former is updated under a hierarchical setting that treats the segment-specific realizations as randomly-drawn samples of a multivariate Gaussian distribution with unknown mean vector and covariance matrix. The primary interest of this model inference and data fusion scheme lies in updating the prior distribution of these hyper-parameters. For the case that the data points are considered to be *i.i.d.* described using Jeffreys prior distributions, the marginalization of the prediction error parameters leads to an explicit formulation. When each segment comprises a sufficiently large number of data points, the model parameters and initial conditions can be inferred from each segment of data using an efficient Laplace asymptotic approximation. In practice, this condition can easily be satisfied since time-history vibrational data often comprise a large number of data points, especially when the sampling rate is large. Furthermore, introducing this approximation allows marginalizing the initial conditions analytically. Once the Gaussian approximations of the model's dynamical parameters are obtained from each segment of data, they are combined to compute the marginal posterior distribution of the hyper-parameters. Thanks to interesting features of Gaussian distributions which allows it to provide explicit formulations for the marginal posterior distribution of the hyper-parameters. As we are interested in the hyper-parameters' MAP estimations, an efficient optimization technique is suggested to search for the most probable values. To enhance the convergence and stability, the objective function is associated with analytical derivatives and initial estimations. By neglecting the uncertainty involved with the hyper-parameters, new formulations are proposed to propagate the uncertainty for predicting response QoI. Subsequently, the posterior predictive distributions of the response QoI are computed analytically and described through the multiplication of student's *t*-distributions. Closed-form formulations are offered for computing the mean and variance of the response QoI. Introducing student's *t*-distributions is novel and beneficial in this context, since they offer more gentle tails as compared to Gaussian predictive distributions suggested in [1,6,7]. In the end, the proposed formulations are outlined in a computational algorithm. The algorithm was tested and verified using numerical and experimental examples. From the results of these examples, the following conclusions and recommendations are made:

- The variability of parameters captured over multiple segments has dominant contribution to the overall uncertainty as compared to the parametric uncertainty obtained from each segment. This dominant impact often exists when the underlying deterministic model is highly-misspecified.

- Dynamical models are often misspecified with respect to ambient conditions, environmental effects, and loading characteristics. Using the proposed hierarchical model is strongly recommended to provide reliable posterior distributions.
- The limiting choice suggested for the prediction error parameters eliminates producing large uncertainties due to imposing subjective prior distributions on the prediction errors. Propagating the parametric uncertainty predicts response QoI with superb accuracy and gives reasonable uncertainty margins.

**Acknowledgement**

Financial support from the Hong Kong Research Grants Council (RGC) under the grants 16234816 and 16212918 is gratefully acknowledged. We also sincerely thank Prof C. C. Chang for sharing the laboratory equipment. This paper is drawn from the PhD dissertation of the first author conducted at the Hong Kong University of Science and Technology and Sharif University of Technology under a joint PhD program. The first author would like to gratefully acknowledge Professor Fayaz Rahimzadeh Rofooie for kind support and supervision at Sharif University of Technology.